\documentclass[3p, twocolumn]{elsarticle}

\usepackage{todonotes} 
  \presetkeys{todonotes}{inline}{}
\usepackage{graphicx}
\usepackage{subcaption}
\usepackage{dcolumn}
\usepackage{bm}
\usepackage{siunitx} 
\usepackage{amsmath}
\usepackage{amssymb}
\usepackage{booktabs}
\usepackage{array}

\usepackage[acronym, nonumberlist]{glossaries}
\makeglossaries
  \newacronym{ADC}{ADC}{Analog-to-Digital Converter}
  \newacronym{CRES}{CRES}{Cyclotron Radiation Emission Spectroscopy}
  \newacronym{DAQ}{DAQ}{Data Acquisition}
  \newacronym{FPGA}{FPGA}{Field-Programmable Gate Array}
  \newacronym{IQ}{IQ}{In-phase / Quadrature}
  \newacronym{SNR}{SNR}{Signal-to-Noise Ratio}
  \newacronym{RF}{RF}{Radio-Frequency}
  \newacronym{IF}{IF}{Intermediate Frequency}
  \newacronym{ROACH2}{ROACH2}{Reconfigurable Open-Architecture Computing Hardware version 2}
  \newacronym{CASPER}{CASPER}{Collaboration for Astronomical Signal Processing and Electronics Research}
  \newacronym{ROI}{ROI}{Region of Interest}
  \newacronym{STFT}{STFT}{Short-Time Fourier-Transform}
  \newacronym{UDP}{UDP}{User Datagram Protocol}
  \newacronym{FIR}{FIR}{Finite Impulse Response}
  \newacronym{IIR}{IIR}{Infinite Impulse Response}
  \newacronym{HDF5}{HDF5}{Hierarchical  Data  Format,  version  5}

\usepackage{hyperref}

\newcommand{\CC}{C\nolinebreak\hspace{-.05em}\raisebox{.4ex}{\tiny\bf +}\nolinebreak\hspace{-.10em}\raisebox{.4ex}{\tiny\bf +}}

\newcommand{\kr}{${}^{83\mathrm{m}}$Kr}

\DeclareSIUnit{\sig}{\ensuremath{\sigma}}

\journal{Nuclear Instrumentation and Methods A}

\begin{document}

\begin{frontmatter}

\title{The real-time data processing and acquisition system for Project 8 Phase II}

\author[10]{A.~Ashtari~Esfahani\fnref{aashtari}}
\author[10]{A.~Banducci}
\author[4]{S.~B\"{o}ser}
\author[7]{N.~Buzinsky\fnref{nbuzinsky}}
\author[10]{R.~Cervantes\fnref{rcervantes}}
\author[4,10]{C.~Claessens\corref{cor}}\ead{claesc@uw.edu}
\author[9]{L.~de~Viveiros}
\author[4]{M.~Fertl}
\author[7]{J.\,A.~Formaggio}
\author[1]{L.~Gladstone\fnref{lgladstone}}
\author[8]{M.~Grando\fnref{mgrando}}
\author[8]{M.~Guigue\fnref{mguigue}}
\author[10]{J.~Hartse}
\author[11]{K.\,M.~Heeger}
\author[8]{A.\,M.~Jones\fnref{mjones}}
\author[6]{K.~Kazkaz}
\author[8]{B.\,H.~LaRoque}
\author[4]{A.~Lindman}
\author[1]{B.~Monreal}
\author[11]{J.\,A.~Nikkel}
\author[10]{E.~Novitski}
\author[8]{N.\,S.~Oblath\corref{cor}}\ead{noah.oblath@pnnl.gov}
\author[3]{W.~Pettus}
\author[10]{R.\,G.\,H.~Robertson}
\author[10]{G.~Rybka}
\author[11]{L.~Salda\~{n}a\fnref{lsaldana}}
\author[11]{P.\,L.~Slocum}
\author[1]{Y.-H.~Sun}
\author[11]{P.\,T.~Surukuchi}
\author[11]{A.\,B.~Telles}
\author[4]{F.\,Thomas}
\author[8]{M.~Thomas\fnref{mthomas}}
\author[5]{T.~Th\"{u}mmler}
\author[6]{L.~Tvrznikova\fnref{ltvrznikova}}
\author[8,10]{B.\,A.~VanDevender}
\author[11]{T.\,E.~Weiss}
\author[2]{A.~Young\fnref{ayoung}}
\author[7]{E.~Zayas}
\author[9]{A.~Ziegler\fnref{aziegler}}

\address[1]{Case Western Reserve University, Cleveland, OH, USA 44106}
\address[2]{Center for Astrophysics | Harvard \& Smithsonian, Cambridge, MA, USA 02138}
\address[3]{Indiana University, Bloomington, IN, USA 47405}
\address[4]{Johannes Gutenberg-Universit\"{a}t, Mainz, Germany 55099}
\address[5]{Karlsruhe Insitute of Technology,Karlsruhe, Germany 76021}
\address[6]{Lawrence Livermore National Laboratory, Livermore, CA, USA 94550}
\address[7]{Massachusetts Institute of Technology, Cambridge, MA, USA 02139}
\address[8]{Pacific Northwest National Laboratory, Richland, WA, USA 99354}
\address[9]{Pennsylvania State University, University Park, PA, USA 16802}
\address[10]{University of Washington, Seattle, WA, USA 98195}
\address[11]{Yale University, New Haven, CT, USA 06520}


\cortext[cor]{Corresponding author}
\fntext[aashtari]{Now at Sharif University of Technology, Tehran, Iran}
\fntext[nbuzinsky]{Now at University of Washington, Seattle, WA, USA 98195}
\fntext[rcervantes]{Now at Fermi National Accelerator Laboratory, Batavia, IL, USA 60510}
\fntext[lgladstone]{Now at Institute for Defense Analyses, Alexandria, VA, USA 22305}
\fntext[mgrando]{Now at Invictus Animus Research and Design, Centennial, CO, USA 80015}
\fntext[mguigue]{Now at Sorbonne Universit\'{e}, Universit\'{e} Paris Cit\'{e}, CNRS/IN2P3, 75005 Paris, France}
\fntext[mjones]{Now at Ozen Engineering, Sunnyvale, CA, USA 94025}
\fntext[lsaldana]{Now at The Est\'{e}e Lauder Companies, New York, NY, USA 10153}
\fntext[mthomas]{Now at Booz Allen Hamilton, San Antonio, TX, USA 78226}
\fntext[ltvrznikova]{Now at Waymo, Mountain View, CA, USA 94043}
\fntext[ayoung]{Now at Radboud University, 6500 GL Nijmegen, The Netherlands}
\fntext[aziegler]{Now at HRL Laboratories, Malibu, CA, USA 90265}

\date{\today}

\begin{abstract}
In Phase II of the Project~8 neutrino mass experiment, electrons from the decays of tritium or \kr{} are detected via their $\approx$\qty{26}{GHz} cyclotron radiation while contained within a circular waveguide.  The signal from a given electron is characterized as a brief chirp, lasting $\lesssim$\qty{10}{ms} and changing in frequency by $\lesssim$\qty{1}{MHz/ms}.  To detect these signals, the Project~8 collaboration developed a \gls{DAQ} system tailored to the signal properties.  The \gls{DAQ} is responsible for simultaneously selecting up to three \qty{100}{MHz}-wide frequency windows to study, detect, and trigger on likely signals from different electron kinetic energies, and for writing the relevant data to disk.  We describe the Phase~II \gls{DAQ} system in detail and address how the system was used for data-taking operations.
\end{abstract}


\end{frontmatter}

\section{Introduction}

The Project~8 experiment aims to measure the neutrino mass using the tritium endpoint method~\cite{Drexlin:2013lha,Formaggio:2021nfz}.  In a \gls{CRES}-based experiment such as Project~8 \cite{Project8:2017nal,PhysRevD.80.051301}, the kinetic energies of electrons are measured by detecting the cyclotron radiation from individual electrons while they are trapped in a magnetic field~\cite{ashtari2019a}.  The cyclotron radiation frequency is inversely proportional to the total energy of the electron:
\begin{equation}
\label{eq:cres}
    f_c = \frac{1}{2\pi}\frac{eB}{m_e + E_{\rm{kin}}/c^2},
\end{equation}
where $e$ is magnitude of the electron charge, $B$ is the magnetic field strength, $m_e$ is the electron rest mass, $E_{\rm{kin}}$ is the electron kinetic energy, and $c$ is the speed of light. In Phase~II\footnote{The Project~8 experiment is divided into four phases~\cite{Project8:2017nal}, with Phase~I being the initial demonstration of \gls{CRES}~\cite{Project8:2014ivu}, Phase~II performing the first \gls{CRES} tritium beta decay endpoint measurement~\cite{phase_II_prl, phase_II_prc}, Phase~III demonstrating the ability to scale up in volume and realize an atomic tritium source~\cite{Project8:Snowmass2022}, and Phase~IV being the final neutrino-mass measurement~\cite{Project8:Snowmass2022}.  } of Project~8 \cite{phase_II_prl,phase_II_prc} the magnetic field strength is $\approx$\qty{1}{T}, putting the frequencies of interest for \kr{} conversion-electron decay and molecular tritium ($\mathrm{T_2}$) beta decay at $\approx$\qty{26}{GHz}.

\begin{figure}
\centering
\includegraphics[width=\linewidth]{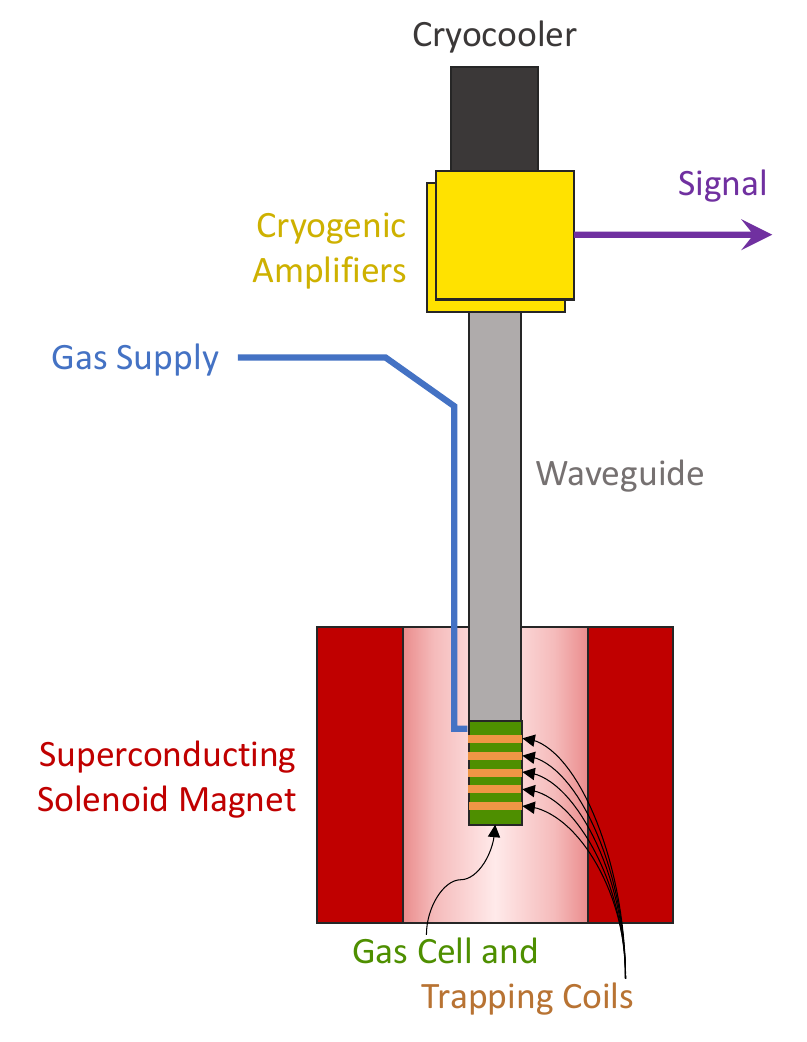}
\caption{In Phase II of Project 8, a waveguide was inserted into the bore of a superconducting magnet. A gas cell near the end of the waveguide is filled with $\mathrm{^{83m}Kr}$ or $\mathrm{T_2}$ gas. The superconducting solenoid magnet provides a background field of \qty{0.959}{\tesla}.  Small copper coils around the cell are powered with $\leq 1\,\si{A}$ to create magnetic-field dips that trap some of the electrons emitted in nuclear decays. The electrons spiral around the magnetic field lines while oscillating up and down in the trapping field and emitting cyclotron radiation. This radiation is detected by an antenna at the far end of the waveguide, amplified by cryogenic amplifiers, and finally filtered and recorded by the \gls{DAQ}.}
\label{fig:phase_2_cartoon}
\end{figure}

The core of the Phase~II experiment~\cite{phase_II_hardware_ref} (\autoref{fig:phase_2_cartoon}) is a circular waveguide that both contains the radioactive gas (it was filled with either \kr{} or T$_{\rm{2}}$) and transmits the emitted cyclotron radiation through its coupling to the waveguide's TE$_{11}$ mode.  Electrons from the source gas are magnetically trapped within the waveguide so they can be observed for an extended period of time.  The magnetic field consists of a \qty{0.959}{\tesla} background field created by the superconducting solenoid magnet, and small ($\mathcal{O}(\qtyrange[range-units = single, range-phrase = -]{
0.05}{0.5}{\milli\tesla})$) dips in the field---the electron traps---created by a set of trapping coils.  All data are collected with a single \gls{RF} antenna at one end of the waveguide.  Analog processing---consisting of amplifying, mixing, and band-pass and low-pass filtering \cite{phase_II_hardware_ref}---reduces the frequencies of interest to the \gls{IF} band near \qty{1}{GHz} (the signal is mixed down by \qty{24.5}{\giga\hertz}). The analog RF stage further sets the \gls{SNR} and total power to match the specifications (e.g. the voltage range) of the digitizer.  In the Phase~II experiment, we need to observe a $\approx$\qty{4}{keV} span around the tritium endpoint to study the backgrounds above the endpoint and the shape of the spectrum below. This translates to a $\approx$\qty{200}{MHz} band in frequency space.

The purpose of the \acrfull{DAQ} system for a \gls{CRES} experiment is to enable the accurate and precise frequency measurement of the particle's cyclotron radiation.  In the context of Phase~II, the \gls{DAQ} system was optimized with a primary focus on efficiently recognizing event candidates and writing the relevant raw data to disk (i.e.~no online reconstruction), and a secondary focus on minimizing the total data volume.  
The Phase~II \gls{DAQ} is able to digitize a single analog input and simultaneously process three frequency intervals of \qty{100}{MHz} each.
The center of each frequency band can be set independently and the system operates without deadtime in either a continuous streaming mode or a triggered mode.
Without triggering the data rate is 600\,\si{MB/s}. Hence, keeping the total recorded data volume for the planned \qty{100}{day} tritium run period below the targeted maximum disk space of $200 \,\si{TB}$ required efficient rejection of noise by the trigger.
Event reconstruction and careful filtering of remaining false events (noise fluctuations mimicking events) are performed offline \cite{ter_paper}.

\begin{figure}
\centering
\includegraphics[width=\linewidth]{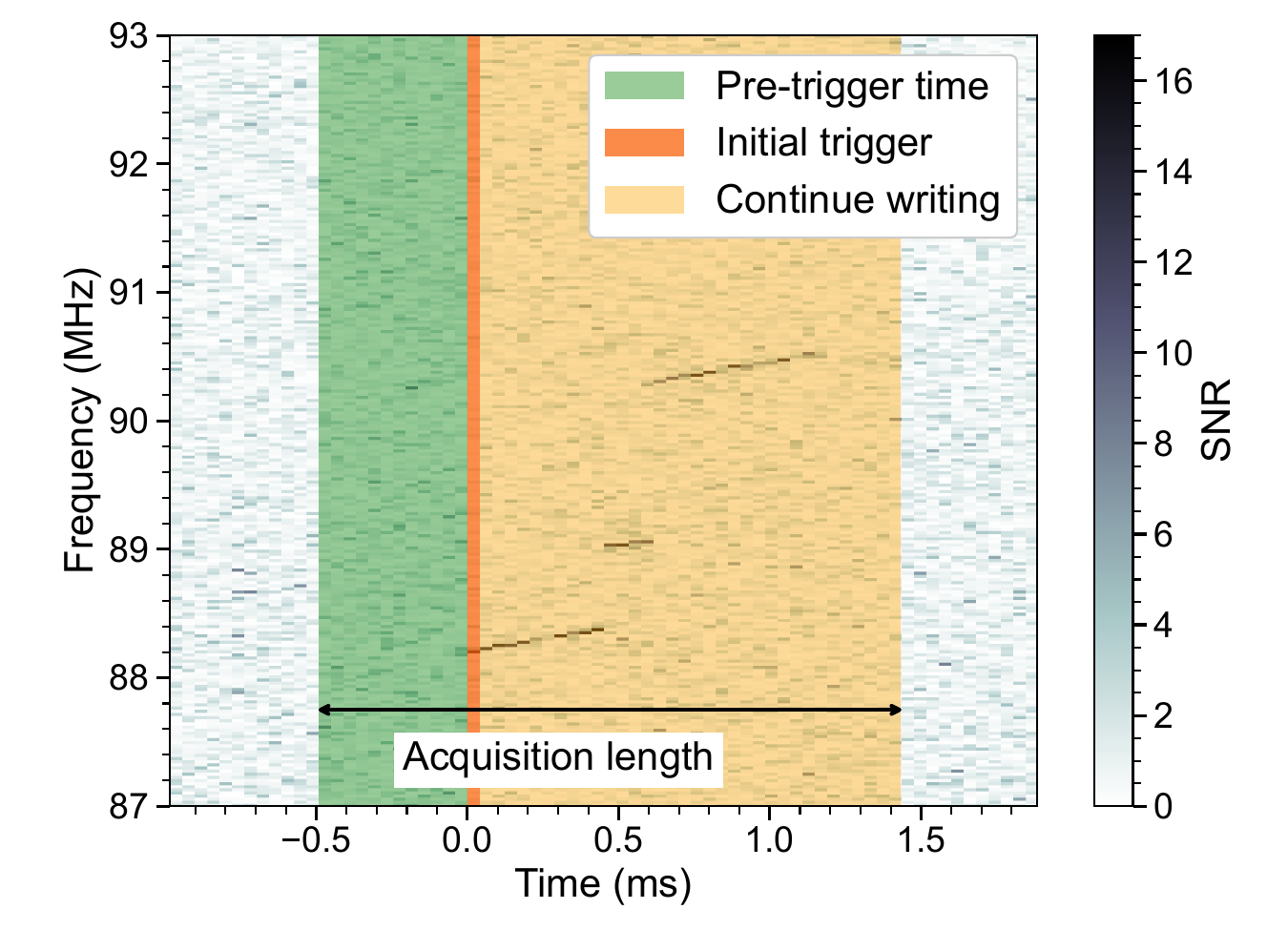}
\caption{Cyclotron radiation from a magnetically trapped electron is observed as a sequence of chirps of varying time lengths and distances in frequency in a spectrogram produced using a short-time Fourier Transform on time-series data down-converted to a range of \qtyrange{0}{100}{MHz}. This spectrogram was produced with the same  \qty{40.96}{\us}-wide rectangular window used in the \gls{DAQ} and data analysis. The goal of the triggered data-taking is to reduce data volume by only recording the color-shaded time span.  Details of how an acquisition is formed are given in \autoref{sec:psyllid_nodes}.  Pre- and post-trigger times here are for illustration only.}
\label{fig:trigger_logic_event_zero}
\end{figure}

\autoref{fig:trigger_logic_event_zero} shows the spectrogram for a typical CRES event from Phase~II, produced by performing a short-time Fourier Transform on the time-series data.  The \gls{SNR} for the event in \qty{24.4}{kHz}-wide frequency bins is plotted as a function of both frequency and time.  The cyclotron radiation from a single electron is detected as a narrow peak in frequency over time, forming an upward-sloping line.  The initial frequency of the signal provides the initial energy of the electron.  The slope is a result of the electron losing power via cyclotron radiation; $f_c$ typically changes by $\approx$\qty{0.3}{MHz/ms} through this mechanism.  Multiple lines are the result of the electron scattering inelastically off residual gas atoms or molecules while remaining magnetically trapped.  These signal characteristics guided the choices made for the \gls{DAQ} system.
We describe the hardware and software details of this system in \autoref{sec:system_description}. In \autoref{sec:triggerin_in_phase_2} the optimization of the trigger for Phase~II data is presented and its performance evaluated. In \autoref{sec:outlook} we discuss the requirements and plans for future \gls{DAQ} systems of \gls{CRES} neutrino-mass experiments. 

\section{System Description}\label{sec:system_description}

\subsection{Overview}

\begin{figure}
\centering
\includegraphics[width=\linewidth]{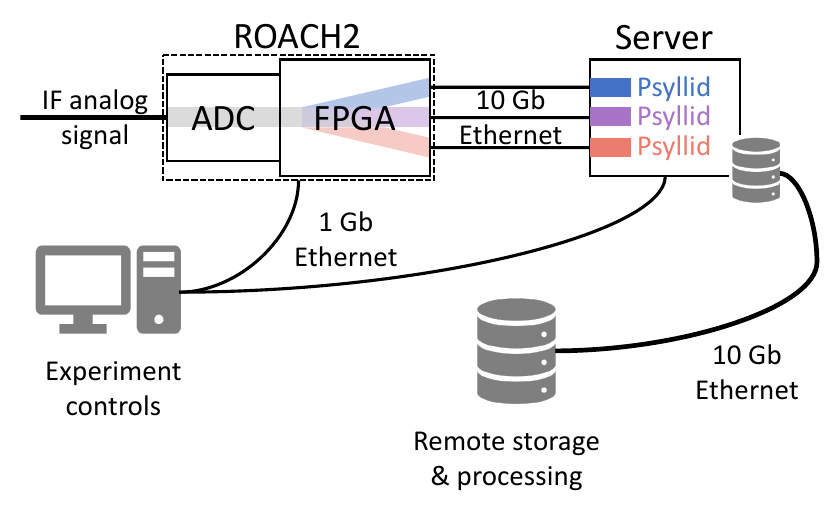}
\caption{Overview of the DAQ system, showing the main hardware components (ROACH2 and server), and the data path from the IF analog signal through digitization, splitting into three channels, data processing, and storage.  Psyllid is the data-processing software described in \autoref{sec:dataprocsoftware}.  The system is connected to the experiment controls over 
 a \qty{1}{\giga\bit} Ethernet network.}
\label{fig:daq_system}
\end{figure}

The \gls{DAQ} system comprises an \gls{ADC} coupled to a \gls{FPGA} that performs initial processing of the data and a data-processing server to analyze the data in real time and write the time series to data files.  \autoref{fig:daq_system} shows the overall \gls{DAQ} system.

The \gls{ADC} and \gls{FPGA} are part of the \gls{ROACH2} platform~\cite{patel2014a, jiang2012a} developed by the \gls{CASPER}~\cite{hickish2016a}. The single channel of digitized data at the IF band is split into three channels on the \gls{FPGA}, each with an independent digital downconversion step.  Each channel is divided into a frequency-domain path and a time-domain path, both of which are transmitted over high-speed Ethernet to the data-processing server. We chose to start the time- and frequency-domain data paths on the \gls{FPGA} to take advantage of its real-time processing capabilities and to simplify the analysis performed by the data-processing server.  The \gls{ROACH2} timing was referenced to an atomic rubidium frequency standard to ensure coherent timing throughout data taking.

Each channel is processed on the server by an independent instance of the data-processing software.  A software trigger uses the frequency-domain data to detect the presence of electron signals because the signals are narrow in frequency and the noise in a single frequency bin is smaller than the total noise present in the time-domain signal.  The trigger algorithm is designed to produce continuous time-domain acquisitions of flexible length with no deadtime, where an ``acquisition'' is defined as a contiguous block of data, either in the time or frequency domain. This way we ensure that we can trigger on events such as the one shown in \autoref{fig:trigger_logic_event_zero}. The time-domain data representing the duration of a triggered acquisition are written to disk to preserve the rawest form of the digitized data for later analysis.  Data are stored temporarily on the server before being transferred to a remote system for processing and long-term storage.

The trigger algorithm is based on the concept of frequency-mask triggering. The in-streaming frequency-domain data are compared to a pre-recorded frequency mask\footnote{Typically, the mask is recorded right before triggered data-taking by averaging over the frequency-domain data for a period of time.} that accounts for any variation in the noise level with frequency. As soon as the mask level is exceeded at any frequency, the algorithm checks if the start-writing condition is fulfilled. This condition can be dynamically configured at runtime and typically consists of a requirement for one or more high-power bins within a certain time interval. To prevent the event start from being missed if only a later part of the event meets the start-writing condition, since the starting frequency is critical for measurig the initial electron energy, a fixed amount of data is buffered and written out as pre-trigger samples. An acquisition is ended if no new occurrence of high power in the frequency domain is detected for a set amount of time.

\subsection{Hardware and Firmware}

The analog RF chain of the Phase~II experiment prepares the signal for digitization~\cite{phase_II_hardware_ref}.  A mixer shifts the signals down to the IF frequency band and a band-pass filter reduces image noise from frequencies above and below the frequency range of interest.  Finally, a combination of an amplifier and a filter set the total power near the optimum level for the digitizer at the sampling rate used and prevents noise aliasing into the first Nyquist zone of digitization.  

The front-end of the DAQ system (\autoref{fig:daq_system}) is the \gls{ROACH2} board, which features an \gls{ADC} implemented on a daughter board~\cite{patel2014a, jiang2012a}, a Xilinx Virtex~6 \gls{FPGA}, and a PowerPC 440EPx processor that runs a Linux operating system.  Control of the \gls{ROACH2} platform is conducted over a \qty{1}{\giga\bit} Ethernet link.  A second daughter board adds four SFP+ (enhanced Small Form-factor Pluggable) \qty{10}{\giga\bit} Ethernet ports to the platform for high-speed digital data transmission.  The data-processing server features an Intel Xeon processor and two dual \qty{10}{\giga\bit} Ethernet network interface cards for receiving the data over a dedicated connection to the \gls{ROACH2} and transmitting it to off-site storage.

The \gls{ADC} daughter board in use has an EV8AQ160 ADC chip that is capable of 8-bit sampling at a rate of up to \qty{5}{GS/s}.  We use a digitization rate of \qty{3.2}{GS/s} to capture a \qty{1600}{MHz}-wide IF band.  

\begin{figure}
\includegraphics[width=\linewidth]{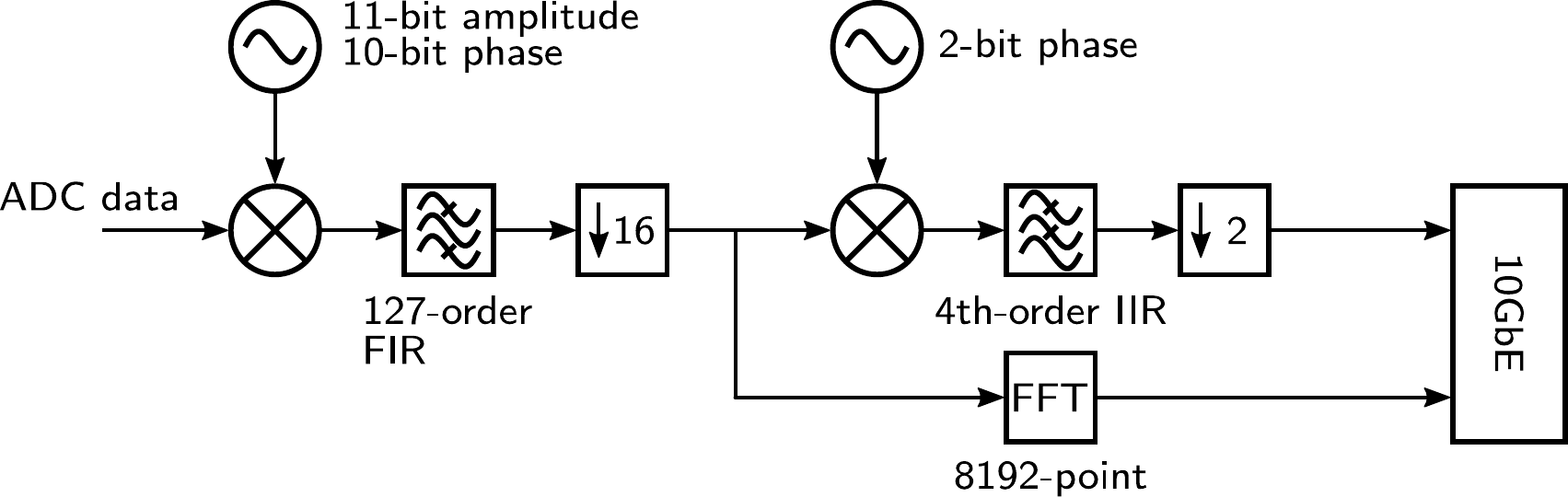}
\caption{Processing implemented on the \gls{FPGA} for each digital channel.  The signal digitized by the ADC is downconverted and its FFT computed.  Both time- and frequency-domain data are streamed over \qty{10}{Gb} Ethernet.}
\label{fig:channeldsp}
\end{figure}

On the \gls{FPGA}, data are processed with a custom firmware configuration.  The single input channel is split into three channels that process data in parallel.  An overview of the processing implemented in each of the three channels on the \gls{FPGA} is shown in \autoref{fig:channeldsp}.  The channels have independently-tunable center frequencies (tuning step size: \qty{3.125}{MHz}) that are used to select \qty{100}{MHz}-wide bands from within the total IF bandwidth. 

When the data from the ADC appear at the input to each channel, they require  decimation by a factor of 16, since the input data are digitized at \qty{3.2}{GS/s} and the \gls{FPGA} is clocked at 200\,MHz. For each channel, the \qty{100}{MHz}-wide band of interest is down-converted with a single-sideband mixer and 128-tap (127th-order) \gls{FIR} band-pass filter. Decimation is achieved by computing every sixteenth filter output to yield a serial stream of time-domain samples at the \gls{FPGA} clock rate. The data are then routed to two parallel paths, one for time-domain processing and one for frequency-domain processing.  On the two paths the data are segmented into 8192-sample ``records'' that are handled simultaneously in parallel.  A record ID value is assigned to guarantee that we can identify which time and frequency records correspond to the same period in time.

The frequency-domain processing computes an 8192-point complex-valued discrete Fourier transform (DFT) using a pipelined fast Fourier transform (FFT) architecture. This yields a spectral resolution of 24.4\,kHz. The negative-frequency samples do not contain any unique information compared with the positive-frequency samples, so only the 4,096 complex-valued spectral samples in the positive half-spectrum are retained.

In parallel, the time-domain processing produces an \acrshort{IQ} (\acrlong{IQ}) sampling of the 100\,MHz wide channel by centering the channel on DC, passing it through a fourth-order \gls{IIR} low-pass filter and downsampling by a factor of two. The \gls{IIR} filter is implemented using a scattered look-ahead architecture, which allows pipelining recursive loops to meet timing constraints while guaranteeing stability~\cite{parhi1989a}.

The output of the frequency-domain and time-domain paths are requantized to 8 bits per sample to compensate for bit growth throughout the processing. The data are then packetized such that frequency-domain samples for a single record are inserted into the payload of one packet, and the time-domain samples that correspond to the same DFT window are inserted into the payload of the following packet. Application layer headers are attached to each packet before being transmitted over 10\,Gb Ethernet. At the transport layer, \gls{UDP} is used.

\subsection{Multi-channel Use Cases}

The multi-channel implementation with independent central frequencies enables the ROACH2 to record data from different frequency ranges simultaneously. This offers many possible applications:

\begin{itemize}

\item Due to the electron motion in the waveguide, the received signal frequencies are Doppler-shifted. This leads to frequency modulation and hence sidebands in the spectrum.  For a large modulation frequency, the separation between the main carrier frequency and sidebands is larger than the bandwidth of a single channel (\qty{100}{MHz}). Using one channel to record the main carriers and the other two channels to record the sidebands would allow us to record all visible bands. The Phase~II experiment was configured in a way that did not result in visible sidebands and this application was never used. 

\item Project~8 uses the \kr{} spectrum for calibration measurements. The energies of the internal-conversion lines with the highest branching ratios in \kr{} decay  are the K line at $\approx$\qty{17.8}{keV}, the L lines at $\approx$\qty{30.4}{keV}, and the M lines at $\approx$\qty{31.9}{keV}. In Phase~II the separations of these lines in the CRES frequency spectrum are $\approx$\qty{600}{MHz} ($\mathrm{L} - \mathrm{K}$ lines) and $\approx$\qty{70}{MHz} ($\mathrm{M} - \mathrm{L}$ lines). After analog mixing with \qty{24.5}{GHz}, all three lines fall into the \qty{1.6}{GHz} bandwidth of the ROACH2 and the three channels can be used to record them simultaneously.

\item 
For the tritium spectrum analysis in Phase~II, the energy region of interest corresponds to a much larger frequency range than the 100\,\si{MHz} bandwidth of a single channel. Therefore, the channels' central frequencies were configured such that the 3 channels were adjacent to each other and together cover the full tritium region of interest (see \autoref{sec:tritium_channel_configuration}). Because the full time-series is recorded for each detected event candidate, splitting the spectrum into 3 frequency ranges reduces the recorded data volume per event by 3 compared to an FPGA implementation with a single 300\,\si{MHz}-wide channel, as most events are seen by only one channel. 

\item For trigger optimization and other systematic studies, it is useful to be able to record the same frequency range with separate channels. For example, to compare the trigger efficiency for different register settings on the ROACH2 or different software trigger configurations, one channel can serve as reference with a constant configuration, while the other two channels are used to explore the configuration parameter space (see \autoref{sec:trigger_parameter_space}).
\end{itemize}

\subsection{Data-Processing Software}\label{sec:dataprocsoftware}

On the data-processing server, the software trigger is implemented as part of the real-time triggering package Psyllid~\cite{psyllid} 
(written in \CC). Its purpose is to receive \gls{UDP} packets via a network port, analyze them in real time (each packet covers \qty{40.96}{\us}) and write data to disk.  The file format used is the ``Egg'' version 3 format \cite{monarch}, a  file  format  that is  based  on  the  industry standard \gls{HDF5} \cite{hdf5}. The Psyllid architecture was designed to:

\begin{itemize}
\item be modular and configurable at runtime,
\item process data in real time with no deadtime,
\item safely run the signal processing steps in parallel, and
\item integrate into an existing Dripline \cite{dripline, laroque_dripline} experiment control system.
\end{itemize}

The remainder of \autoref{sec:dataprocsoftware} describes the components of the data-processing software in detail.

\subsubsection{Software Framework: Midge}

Psyllid is based on the Midge framework~\cite{midge,furse}. The working principle behind Midge is that data are produced, read, and processed by different kinds of operators called ``nodes''.  A ``data stream'' is used to connect a node writing data to a node reading data.  The different types of nodes are:
\begin{itemize}
\item Producers: nodes that only write output data to a data stream(s), and do not read input data from a data stream;
\item Transformers: nodes that read input data from a data stream(s), and write output data to a data stream(s);
\item Consumers: nodes that only read data from a data stream(s), and do not write data to a data stream.
\end{itemize}

Data streams are circular buffers of data objects, which make it possible to continuously process data as long as data are read from the buffer as quickly, on average, as it is written to the buffer.  Each data stream carries a single type of data.  Producers and transformers (writing) move along the data stream writing to each data object.  Consumers and transformers (reading) follow the writing operator along the data stream, reading from each data object.  Each node runs in its own thread which allows for parallel processing. Mutexes are used to avoid collisions between reading and writing nodes, and since a writing node cannot pass a reading node, nodes are prevented from overwriting data that has not yet been read.

In principle, nodes can use an arbitrary number of input and output streams, though each data stream has a single producer. All nodes can be connected via data streams in almost arbitrary ways, where each operator takes action on the incoming data as soon as it ``appears'' on the output of the upstream operator. A concrete example of a node configuration used in Phase~II is shown in \autoref{fig:psyllid_triggered_mode}. Upon initialization of a Midge instance, the network of nodes is established and connected according to a pre-configured structure.

\subsubsection{Data processing overview}

In Psyllid we implemented a feature that allows rapid starting and stopping of data processing.  Prior to initializing the Midge nodes, Psyllid is in the ``idle'' state.  Psyllid is put in the ``activated'' state before starting data processing---which involves initializing the desired Midge nodes, connecting them via their input and output streams, and starting the threads for each node---so that data processing can start almost instantly when the run is started.  One or more of the nodes must have the ability to pause its data processing, in a way that halts the overall data processing. 
The act of starting a data-taking run puts Psyllid in the ``running'' state by un-pausing the relevant node(s).  The end of the run pauses the node(s) again and returns Psyllid to the ``activated'' state.  In this way, data-taking runs can be started and stopped with essentially no delay in the action of the DAQ system.

For Phase~II of Project~8, we built a number of nodes in Psyllid specifically for processing CRES data.  Pausing and un-pausing is performed by the ``ROACH receiver'' (described below in more detail).  Data taking is performed in two modes: streaming and triggered.  In streaming mode, the time series data received from the ROACH2 is written directly to disk.  In triggered mode, time-series data are written to disk only when an event is present.  In both configurations, the data are processed by Psyllid in three stages:

\begin{enumerate}
    \item Receiving: network packets arrive on the network interface, are transferred into memory, and then are split into time and frequency streams;
    \item Analyzing: the time and frequency streams are processed by different sets of transformer nodes depending on whether Psyllid is in streaming or triggered mode;
    \item Writing: time series data are written to disk.
\end{enumerate}

In all configurations, deadtime is avoided by carefully balancing the workload of the nodes in use.  Each node must process data at least as quickly as the data are received by Psyllid.  As long as the rate of packets being received by Psyllid is the limiting rate in the system, all downstream nodes will stay synchronized with the incoming data.  The consequence of a slower downstream node would be for the buffers of the nodes upstream of the slow node to fill up until eventually arriving packets are dropped when they cannot be written to the packet receiver's output buffer.  In development of the Psyllid configuration used to record Phase~II data, the processing rates of all nodes were checked to ensure that deadtime would not be introduced by Psyllid.

In sections~\ref{sec:streaming_mode} and~\ref{sec:triggered_mode}, we will briefly describe how Psyllid applies the three stages above to operations in streaming and triggered modes, respectively.  Section~\ref{sec:psyllid_nodes} will cover the setup and nodes of Psyllid for production data-taking in triggered mode in more detail.  

\subsubsection{Streaming mode}\label{sec:streaming_mode}
In streaming mode the goal is to write as much data as possible to disk, without introducing deadtime.  

\paragraph{Receiving} The process of receiving data from the ROACH2 involves two nodes: the ``packet receiver'' and the ``ROACH receiver''.  The packet receiver is responsible for taking packets directly from the server's network interface, and copying the data as a raw sequence of bytes to a raw-data object in the packet receiver's output data stream.  

The ROACH receiver takes the raw-data objects from the packet receiver and interprets them as either time- or frequency-domain packets, outputting the two types of data on separate output streams. Each packet contains a single record holding two arrays (\gls{IQ} samples) with 4096 8-bit integer numbers.

\paragraph{Analyzing} No analysis is performed on the data in streaming mode; the frequency-domain data are ignored with a ``terminator'' node, and the time-domain data are passed directly from the ROACH receiver to the writer node.

\paragraph{Writing} The purpose of streaming mode is to write all time-domain data received to disk.  The ``streaming writer'' copies the time-series data directly from each data object into an Egg file.  The Egg file contains a single acquisition covering the entire run period, except in the rare cases when a data-processing problem induces deadtime and starts a new acquisition.

\subsubsection{Triggered mode}\label{sec:triggered_mode}
Triggered mode differs from streaming mode in that we seek to only write data that includes a potential event.  To accomplish this we perform a real-time analysis of the frequency-spectrum data, and then write to disk the time-series data where the signal is present.

\paragraph{Receiving} The process of receiving data in triggered mode is the same as in streaming mode, as described above.

\paragraph{Analyzing} Only the frequency-domain data are analyzed to determine if an event is present; the time-domain data are directly passed from the ROACH receiver to the writer node. The first node in the analysis process is the frequency-mask trigger (FMT), where a configurable frequency-dependent threshold is applied to each record to determine if it contains any signal power.  The output of the FMT is a trigger flag for each record that indicates whether the trigger conditions have been met.

The next node, the event builder, implements logic to set the trigger flag of a record based on the surrounding records and create ``events,'' sets of records likely to contain a full electron signal (or, in principle, that of multiple simultaneous electrons).  Its input is trigger flags from the frequency-mask trigger node, and its output is also trigger flags (not the same as the input), which are read by the writer node.  The event builder uses circular buffers to store a certain number of trigger flags to build events in two ways: (1) a set number of records called a pre-trigger are added before the first triggered record in an event, and (2) a gap between triggered records is filled in if it is smaller than a specified ``skip tolerance.''  

Two further options allowed customization of the triggering process to help reduce noise fluctuations being interpreted as events.  The FMT included an option to use two threshold levels.  The trigger flags would indicate whether the low or both the high and low thresholds were crossed.  When this feature was used, the event builder would require the high threshold to be crossed to trigger an event, but trigger flags crossing the low threshold would define a wider time window for the event, accounting for variability in the event power while maintaining high noise rejection and still seeing the whole event.  The event builder could also require more than one trigger flag before creating an event.  This improved noise rejection since false events due to noise were most likely to be short.  The settings used for all configurable parameters are optimized in \autoref{sec:trigger_optimization}.

\paragraph{Writing} The ``triggered writer'' node has two inputs: time-domain records and trigger flags.  For each record, if a trigger flag is \emph{true}, then the time-domain data for that record is written to an Egg file.  In the end, the Egg file contains a set of acquisitions, each nominally containing the signal from a single event.







\begin{figure}
\centering
\includegraphics[width=0.7\linewidth, trim={0 0cm 0 0},clip]{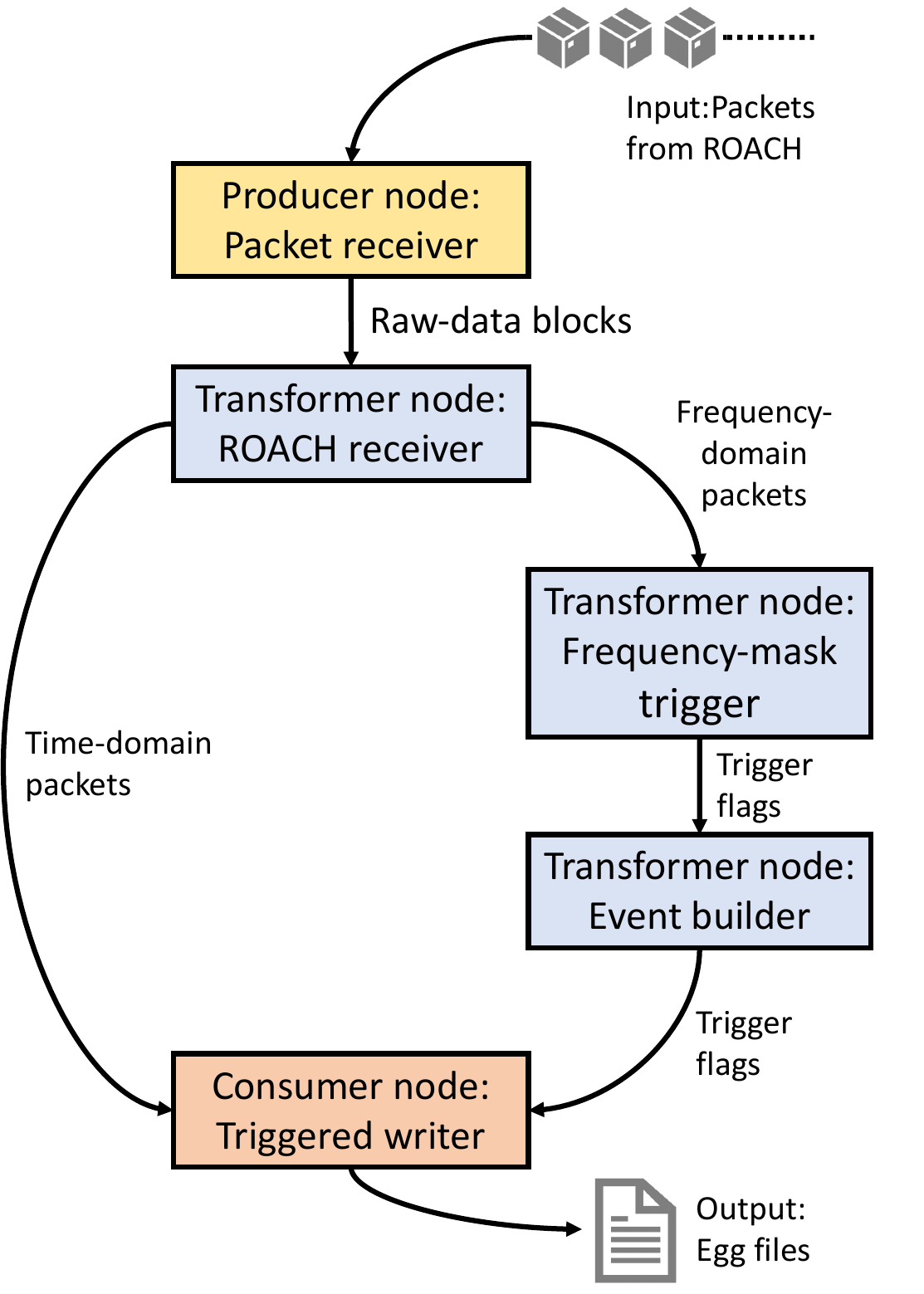}
\caption{Psyllid node configuration in triggered mode. The ROACH receiver interprets packets as either time- or frequency-domain data, and sends the data on the appropriate stream.  Records from the time-domain packets are passed directly to the writer, while their frequency-domain equivalents are inspected for high-power bins by the frequency mask trigger node (FMT). The event builder determines which record IDs comprise the next acquisition. The writer node then writes the time-domain records sorted as acquisitions to an Egg file using the Monarch3~\cite{monarch} library.}
\label{fig:psyllid_triggered_mode}
\end{figure}

\subsubsection{Psyllid Nodes}\label{sec:psyllid_nodes}

The nodes used in triggered mode are depicted in \autoref{fig:psyllid_triggered_mode}. Their functionality is as follows.

\paragraph{Packet receiver}
Psyllid has two types of packet receivers, both of which are producer nodes: the socket receiver and the fast-packet receiver. The socket receiver gets the UDP packets via the standard POSIX socket interface~\cite{sockets}. As this is a simple high level process, it is very useful for testing purposes or for receiving packets from a packet simulator that produces UDP packets at a slower rate. For production data-taking, this interface is too slow because the Linux socket interface uses one system call to capture a single packet. The fast-packet-receiver, on the other hand, gets the packets via the memory-mapped ring buffer~\cite{mmap}. The memory-mapped buffer provides a circular buffer of configurable size that is shared between the kernel and the user. Instead of issuing a system call, the user can read packets from the shared buffer. Both receivers are Midge producers with a single output stream and they both write the packet content as raw blocks of memory (i.e. without interpreting as any particular data structure) to improve data throughput.

\paragraph{ROACH receiver}

The ROACH receiver is a transformer node responsible for dividing the time and frequency data packets onto separate paths.  Its input is a stream of raw data packets from the packet receiver, and it has two output streams, one for each domain of data.  The ROACH receiver is also responsible for implementing the ``pausing'' behavior mentioned above.  In the paused mode, data objects continuously arrive from the packet receiver but are ignored until data processing is un-paused.  In that way, there is no backlog of data from the network interface, and no extra buffering of packets is needed to start a run immediately by un-pausing.

\paragraph{Frequency-mask trigger}

The FMT is a transformer node with one in-stream and one out-stream. The input data type is freq-data and the output data type is trigger-flags. The FMT can run in two different modes, ``add-to-mask'' and ``triggering.'' 

\begin{figure}
\centering
\includegraphics[width=\linewidth]{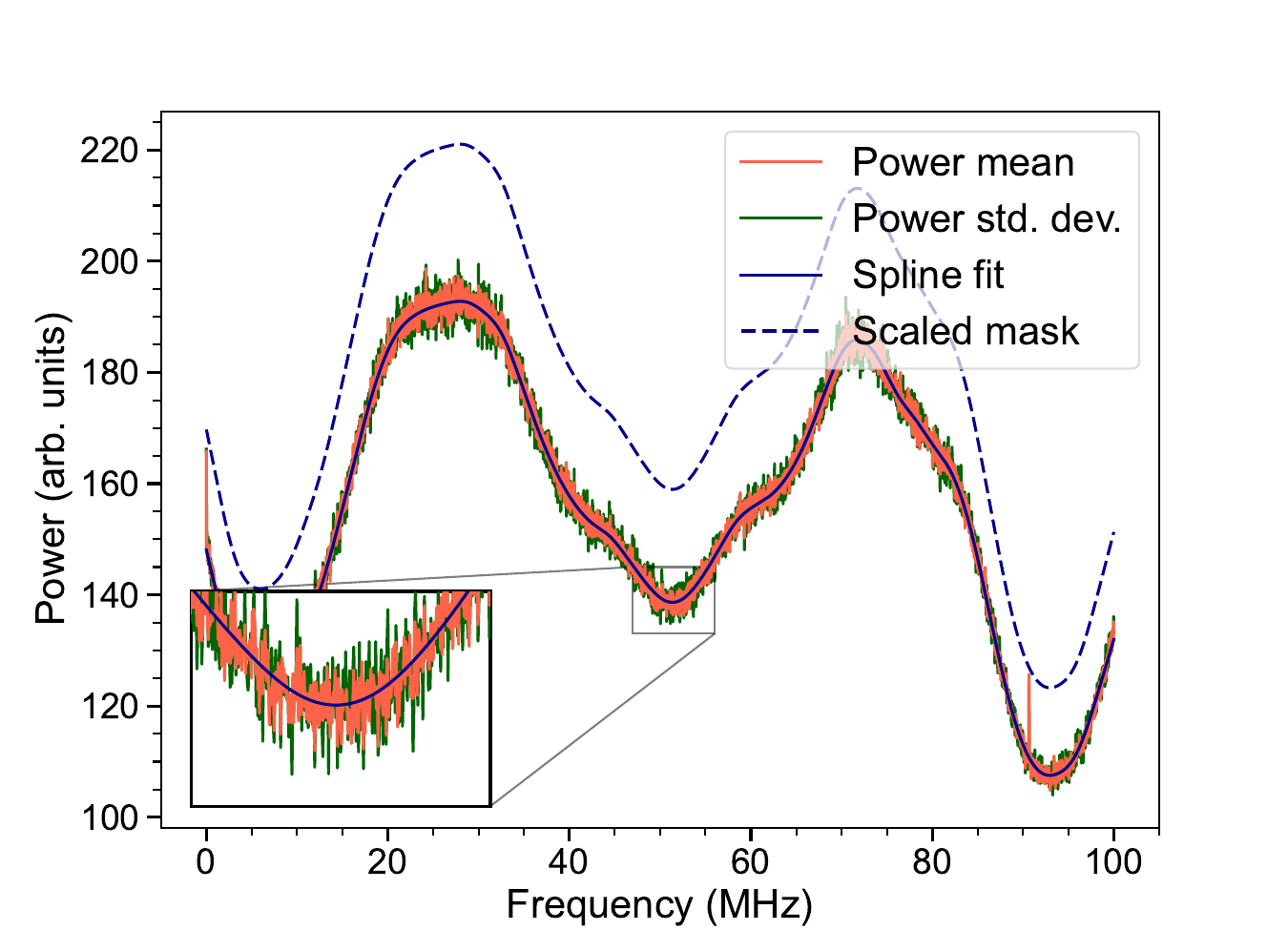}
\caption{Example of a trigger mask calculated from $N=12,000$ frequency-spectrum packets.  The bin-wise mean and standard deviation are shown along with the spline fit and the scaled mask.  The mask threshold in this example is $12*\sigma_{\rm{power}}$.  The inset shows that the spline fit does not under- or over-estimate the shape of the average spectrum even around one of the sharper spectral features.} 
\label{fig:mask_example}
\end{figure}

In the {add-to-mask} mode, the FMT calculates the mean and the variance of the incoming power spectra over time. To do so, it takes the incoming complex-valued array from each packet and sums the absolute square (proportional to power) and the square of the absolute square (variance of the power) separately for each frequency bin. Once a pre-configured number of records, $N$, has been processed, the sums are divided by $N$ and a third-order spline is fitted to the mean and variance. Depending on configurable options, the FMT then either calculates one or two trigger masks equal to $r_{\rm{low}}$ and $r_{\rm{high}}$-times the fitted average background (\autoref{fig:mask_example}) or $r_{\rm{low}}$ and $r_{\rm{high}}$ standard deviations above the background, where $r_{\rm{low}}$ and $r_{\rm{high}}$ are configurable parameters used to calculate the trigger masks. For all of our production data-taking we used the two-threshold option, calculating those thresholds using the standard deviations.  The output-data stream is not used in this mode.



In triggered mode the FMT loops over all bins in the frequency domain data, calculates the power from IQ amplitudes in each bin, and compares the result to its stored trigger masks. 
For each record, a trigger-flag data object is written to the node's output stream.  If the trigger conditions are met, the flag is set to \textit{true}, and otherwise it is set to \textit{false}.

While using a single trigger mask, each bin is compared to the trigger mask until one is found that exceeds the threshold.  When that occurs, the output-data trigger flag is set to \textit{true}, and the loop over bins is stopped.  

While using two masks with different thresholds ($r_{\rm{low}} < r_{\rm{high}}$), the FMT scans through the bins of the frequency spectrum comparing to both thresholds. If a bin crosses the high threshold, then both the trigger flag and the high-threshold flag in the output data are set to \textit{true} and the scan over the bins stops.  If a bin only crosses the low threshold, then the trigger flag in the output data is set to \textit{true} and the scan continues.  When the scan over bins completes, if only the low threshold has been crossed, then only the trigger flag is set to \textit{true} in the output data.


\paragraph{Event builder}

\begin{figure*}[tb]
\centering  
\includegraphics[width=0.7\linewidth]{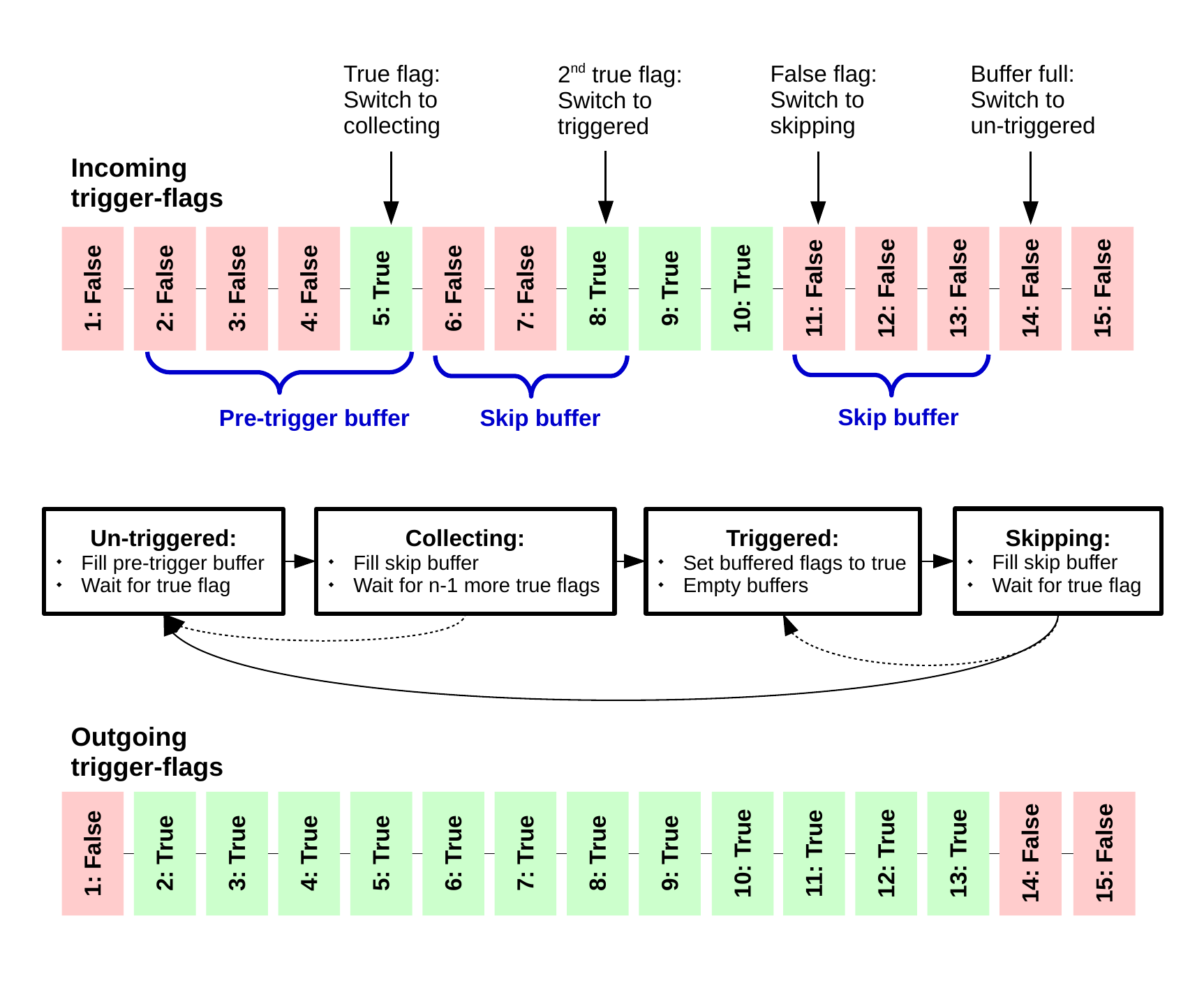}
\caption{Example for trigger-flag processing in the event builder following the FMT, operating in single-threshold mode with $\textnormal{n-triggers}=2$. While in the un-triggered state, in-streaming trigger flags are constantly placed into the pre-trigger buffer. When a new trigger flag arrives and the pre-trigger buffer is full, the first trigger flag in the buffer is moved to the output stream. Once a trigger flag set to \textit{true} arrives (trigger-flag 5), the event builder switches to the collecting state and places incoming flags into the skip buffer. If a new trigger-flag set to \textit{true} arrives before the skip buffer is full, the event builder switches to triggered state. In this state all trigger-flags in the pre-trigger buffer (2 to 5) and in the skip buffer (6 to 8) are labeled as \textit{true} and are moved to the event builder’s output stream. The next two trigger flags (9 and 10) are true because the data are still above threshold and are transferred directly to the event builder's output stream.  When the next incoming trigger-flag is labeled \textit{false} (11), the event builder switches to the skipping state and starts filling the skip buffer again. If no new \textit{true}-flag arrives before the skip buffer is full, the event builder switches back to the un-triggered state. In this configuration the skip-buffer is smaller than the pre-trigger buffer and all flags in the skip buffer are labeled as \textit{true} and moved to the output stream. All records corresponding to the trigger-flags that are moved to the out-stream as true (2 to 13) will be written to an acquisition by the triggered writer. If no flag had been labeled \textit{true} in the collecting state, or a single \textit{true}-flag had arrived in the skipping state, the event builder would have returned to the previous state (as indicated by the dashed arrows). }
\label{fig:psyllid_event_builder_example}
\end{figure*}

The event builder is a transformer node responsible for taking the set of records passing the trigger threshold(s) and adding other records around them to form an acquisition that is likely to capture a full event (as shown in \autoref{fig:psyllid_event_builder_example}).  An acquisition includes a number of records before the acquisition is triggered, called the pre-trigger, and records to fill in small gaps between triggered records, called the skip tolerance.  Both the input and output data types to the event builder are trigger-flags.

The event builder is implemented as a finite state machine with four states: un-triggered, collecting-triggers, triggered, and skipping.  The behavior of the event builder is controlled with three parameters: the pre-trigger size, the skip-tolerance size, and the trigger-requirement size, referred to as ``n-triggers'' in the following sections.  The parameters are configurable at run time, and their use during Phase~II operations is discussed below.

The builder uses two circular buffers to temporarily store trigger flags: the pre-trigger buffer and the skip buffer.  The sizes of those buffers are set by the first two parameters.  The third parameter controls whether multiple records with trigger flags set to \textit{true} are required to start an acquisition.  Trigger-flag data objects on the input stream are sequentially put into one of the two buffers, depending on the state of the builder.  The builder uses the contents of the buffers to determine whether to switch to a new state or remain in the current state.  Trigger-flag data objects are sequentially removed from the end of the buffers (again, depending on the state of the builder) and put into the output data stream with the trigger flag set according to whether an acquisition is triggered.  The next four paragraphs describe the state-machine logic in more detail.


The event builder starts in the un-triggered state.  Input trigger flags are sequentially placed into the pre-trigger buffer.  Once the buffer is full, and as data continues to arrive, trigger flags from the end of the buffer are transferred to the output data stream with their trigger-flag variables set to \textit{false}.  

An event will be triggered by the arrival of n-triggers trigger flags that are set to \textit{true} (requiring the high-threshold flag if using the two-threshold option).  If $\textnormal{n-triggers}=1$, the state switches to triggered.  If $\textnormal{n-triggers}>1$, the state switches to collecting triggers and starts filling the skip buffer.  
Note that the pre-trigger buffer does not need to have filled before trigger flags are checked. This avoids deadtime between events or at the beginning of a run.
In the collecting-triggers state, as further input data arrive, once the number of \textit{true} trigger flags in the skip buffer is equal to n-triggers, the state is switched to triggered.   
If not enough \textit{true} trigger flags arrive while in the collecting-triggers state to meet the trigger-requirement size by the time the skip-buffer is full, the state is switched back to un-triggered. In this case trigger flags are transferred from the pre-trigger buffer to the output data stream to make enough room in the buffer to fit all trigger flags that are currently in the skip-buffer. The transferred flags are set to \textit{false} and no acquisition is written.

When the state switches to triggered, all buffered trigger flags are moved to the output data stream with their trigger flag variables set to \textit{true} (just the pre-trigger buffer if $\textnormal{n-triggers}=1$, and the pre-trigger and skip buffers if $\textnormal{n-triggers}>1$).
Trigger flags set to \textit{true} that arrive at the builder while it is in the triggered state are immediately moved to the output stream. 

The event builder uses its skipping state to handle gaps (e.g. due to noise fluctuations) in the middle of an acquisition.  While in the triggered state, if a \textit{false} trigger flag arrives, the event builder switches to the skipping state, and incoming trigger flags are directed to the skip buffer.  If a \textit{true} trigger flag arrives before the skip buffer is filled, then the event builder switches back to the triggered mode; all of the trigger flags in the skip buffer are moved to the builder's output stream with their trigger-flag variables set to \textit{true}.  Alternatively, if no \textit{true} flags are found before the skip buffer fills, then the acquisition is ended.  Some or all of the trigger flags in the skip buffer are used to add a post-event margin to the acquisition: if the skip buffer size is larger than the pre-trigger buffer size, only the excess flags above the pre-trigger buffer size are moved to the builder's output stream with their trigger-flag variables set to \textit{true} and the remaining flags are transferred to the pre-trigger buffer; otherwise all flags in the skip buffer are moved to the builder's output stream with their trigger-flag variables set to \textit{true} and the pre-trigger buffer remains empty. 
 \autoref{fig:psyllid_event_builder_example} shows an example of a sequence of incoming trigger-flags and indicates into which buffers they would be placed by the event builder.  \autoref{fig:trigger_logic_event_zero} shows how the aspects of the event builder are applied to an actual event.

\paragraph{Writers}

Two consumer nodes are responsible for writing data to disk: the streaming writer and the triggered writer.  Both write data in the Egg file format, which is based on the HDF5 file format~\cite{hdf5}, using the Monarch software library~\cite{monarch}.  The writers also both use the concept of an ``acquisition'' in the Egg standard: a single acquisition represents a contiguously recorded signal.

The streaming writer is a consumer node with one input: time-series data.  It simply writes every record it receives into the Egg file.  If a break in the sequence of record ID numbers is detected, it starts a new acquisition after the break.

The triggered writer is a consumer node with two inputs: time-series data and trigger-flag data.  If the ID values for the time-series data object and trigger-flag data object do not match, the run is stopped so that the data streams can be synchronized. If the trigger-flag is \textit{true}, then the time-series data are written to the Egg file; if it is \textit{false}, then the time-series data are skipped.  Every contiguous block of \textit{true} trigger flags forms a single acquisition.

Psyllid includes a run-time configurable parameter to limit file sizes and write to multiple files sequentially in a single run.  Psyllid guarantees that there is always an Egg file available to write data to without delaying the process of acquiring data.    If file writing had to be paused to close a full file and then open a new one, the writer would not be able to catch up with the backlog of data, resulting in gaps in the data written to disk.  Instead, separate threads are used to open and close files in parallel with the processing and writing of data to disk.  A new file is opened while an existing file is being written to, and once the existing file is full, it can be quickly swapped with the new file so that the writing of data can continue.  The existing file is then closed in parallel with data being written to the new file.  

Unfortunately this practice revealed an issue with our use of the underlying HDF5 library: the C++ interface for HDF5 is not guaranteed to be thread-safe, which caused instabilities in this multi-threaded use of Psyllid.  Psyllid would crash occasionally as a result, and the affected data were not used.  No evidence of corruption to the data recorded while Psyllid ran stably was found.  While this problem was handled for Phase~II operations by detecting software crashes quickly, it will have to be addressed for future use of Psyllid with Egg files.

The decision to start writing to a new file is made after each record is recorded.  As a consequence, triggered events are sometimes split between files.  The first part of a triggered event, which is written to the end of a given file, will result in an event that is included in the analyzed data set.  The second part of a triggered event, which is written to the beginning of the next file, will not result in an event included in the analyzed data set because of the cut on events starting in the first \qty{0.5}{\milli\second} of an acquisition.  
This occurs for only a small fraction of events because the maximum file size is large compared to the size of a single event in the file.
Therefore, while this affects the distributions of track and event characteristics in principle, no effect was observed. 

\paragraph{Terminators}

Two nodes were implemented for stopping a data stream without further processing: the time and frequency terminators.  These consumer nodes were capable of reading data on a time or frequency data stream, respectively, and simply taking no action.  They would therefore ensure that the buffer for the relevant data stream would not fill up.  Only the frequency terminator was used in Phase~II of Project~8, when operating in streaming mode.


\section{Trigger optimization and performance in Project~8 Phase~II}
\label{sec:triggerin_in_phase_2}

A trigger for CRES events has to fulfill a number of  requirements.  The trigger should be highly efficient at identifying reconstructable events and identifying undesired noise as events should be minimized.  Deadtime should be minimized as events can occur at any point in time. 
If the detection efficiency (defined as the fraction of  events that are detected) depends on frequency, that dependence becomes a critical systematic effect in the shape analysis of a CRES spectrum. Therefore, the trigger efficiency should be independent of frequency or at least the frequency dependence has to be very well understood. In the following sections the performance of the ROACH2-Psyllid triggering system is evaluated and the configurable parameters are optimized for the Phase~II running conditions. The trigger efficiency with optimized parameters is studied by applying the trigger to simulated data, and a channel central-frequency configuration is found that minimizes the impact of the trigger efficiency on the Phase II spectrum analysis.


\subsection{Deadtime}

\begin{figure}
\centering
\includegraphics[width=\linewidth]{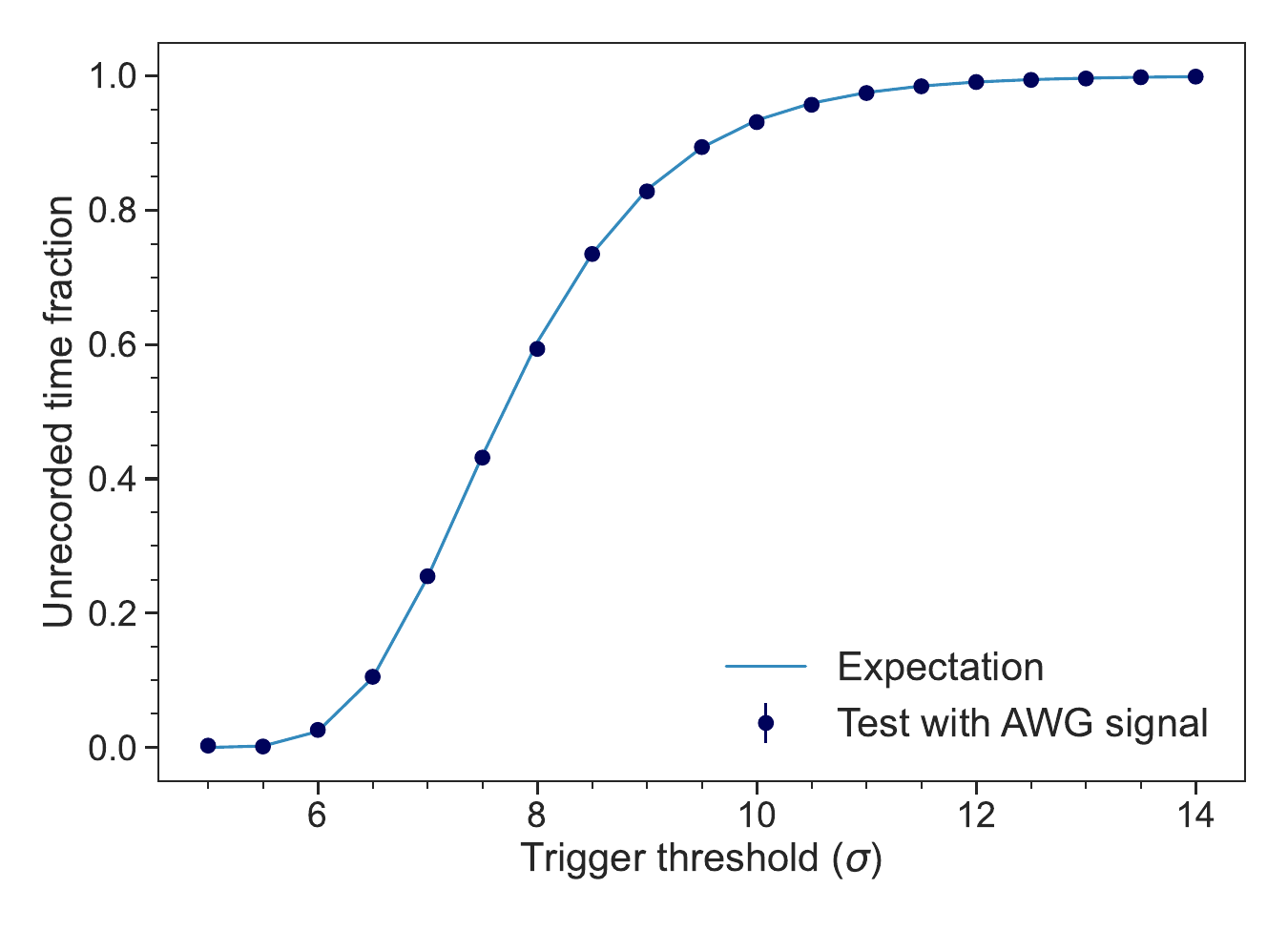}
\caption{Fraction of unrecorded time in a trigger threshold scan. \gls{IQ} white noise from an arbitrary waveform generator (AWG) was injected into the ROACH2 input. Psyllid was configured to write out all records that exceeded the trigger threshold (if at least 1 sample in the record was above the threshold). The recorded time fraction follows the expectation for exponentially distributed noise power in 4096 frequency bins per record.}
\label{fig:AWG_threshold_scan}
\end{figure}

To test if the trigger can operate without deadtime in the production configuration (triggered mode), an Arbitrary Waveform Generator was used to play an \gls{IQ} white noise signal to a ROACH2 that streamed the processed signal in packets to a running Psyllid instance. The Psyllid nodes were configured as described in \autoref{sec:triggered_mode} and a FMT threshold scan was conducted.
The pre-trigger and skip-tolerance buffer sizes were set to 0, so that for each FMT trigger (i.e. after the FMT, without further event building), an acquisition with a single record would be written to file. The fraction of records that were not written to an Egg file is shown in \autoref{fig:AWG_threshold_scan}. A trigger deadtime would lead to an increase of the unrecorded time fraction compared to the theoretical expectation. No large increases were observed; however, at the lowest trigger threshold of $5\sigma$, the test run resulted in a recorded time fraction of 99.72\% instead of the expected $>99.99\%$. One must conclude that in this extreme test case, in which every record is a new trigger, Psyllid reaches its limitations in deadtimeless operation. During Phase~II data-taking, we used trigger thresholds of $\geq 13 \sigma$. No excess of unrecorded time is observed in this range during the deadtime test. 
We conclude that Psyllid operates effectively without deadtime (to the precision of this test) at the trigger rates relevant in Phase~II.



\subsection{Parameter optimization for CRES triggering}
\label{sec:trigger_optimization}

We optimized the trigger parameters according to event properties, such as event duration and SNR, in the Phase II experiment. 
Based on the available disk space in Phase~II and the target to take tritium data for \qty{100}{days}, the goal was set to achieve a recording rate that does not exceed 
\qty{2}{TB/day} for all channels. Since in streaming mode, the 3 channels together process \qty{600}{MB/s}, this corresponds to a data reduction of \qty{96}{\percent}. 

The trigger logic has a total of five configurable parameters as described in \autoref{sec:psyllid_nodes}:
\begin{itemize}
\item pre-trigger time
\item skip-tolerance
\item n-triggers
\item FMT threshold ($r_{\rm{low}}$)
\item FMT high-threshold ($r_{\rm{high}}$)
\end{itemize}
The best compromise between data reduction and detection efficiency for Phase~II production tritium data-taking was achieved with $\mathrm{pre\text{-}trigger~time}=2\,\si{ms}$, $\mathrm{skip\text{-}tolerance}=0.5\,\si{ms}$, $\mathrm{n\text{-}triggers}=2$, $\mathrm{FMT~high\text{-}threshold}=13.5\,\sigma$, and $\mathrm{FMT~threshold}=11\,\sigma$. The optimization procedure is described in the following sections.



\subsubsection{Trigger parameter space}
\label{sec:trigger_parameter_space}


The optimum pre-trigger time is determined by the average event duration ($\approx$\qty{1}{ms}), which depends on the gas pressure and composition in the waveguide, as well as on the trap configuration. In particular, deeper traps lead to longer trapping times and hence longer events. The pre-trigger time is set such that \qty{>90}{\percent} of all event starts would still be contained in the acquisition window even if only the very last bin in the events triggered the acquisition. 
The skip tolerance was scanned between \qty{0.5}{ms} (12~records) and \qty{2.5}{ms} (61 ~records), which roughly corresponds to the range of detected track durations under Phase~II running conditions.

\begin{figure}
\centering
\includegraphics[width=\linewidth]{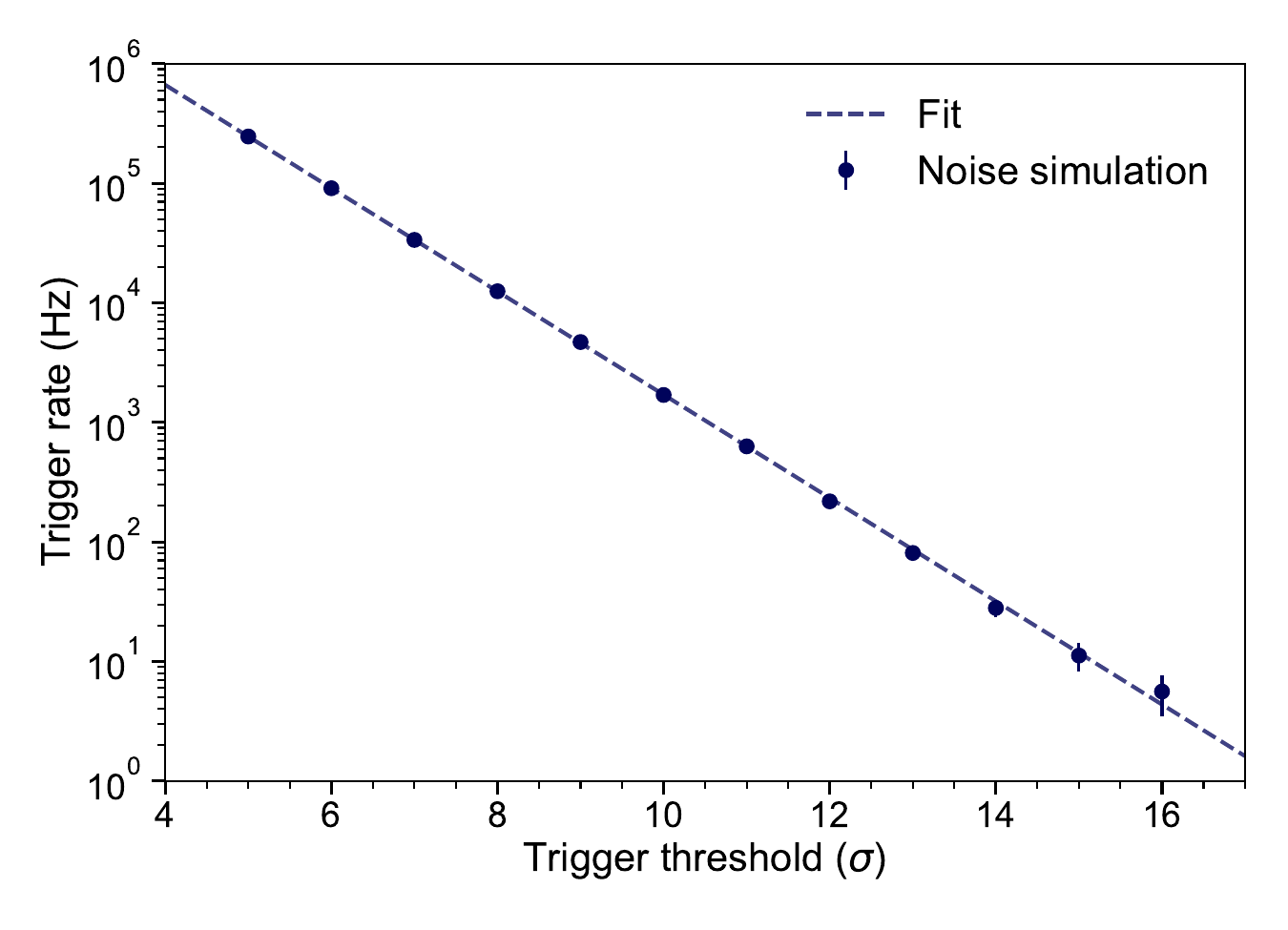}
\caption[Simulated noise trigger rate vs. trigger threshold]{Noise trigger rate vs. FMT threshold for $\textnormal{n-triggers}=1$ in simulated data. Since the noise power is exponentially distributed, the noise trigger rate follows an exponential function.}
\label{fig:trigger_rate_in_exp_noise}
\end{figure}

To pre-select a range for the high-threshold parameter, \autoref{fig:trigger_rate_in_exp_noise} shows the FMT-trigger rate as a function of a set \unit{\sig} threshold for simulated noise. 
Note that the optimization task is not to reduce the noise trigger rate, but instead to limit the filled disk volume per time. Therefore the threshold choices depend on the acquisition lengths. With the assumption that the majority of triggers are noise triggers with no re-triggers (i.e. consecutive triggers within the skip-tolerance), the minimum acquisition length is calculated as the sum of the pre-trigger time plus the minimum time that is recorded after a trigger (skip-buffer length). For a skip-tolerance of \qty{0.5}{ms} and a pre-trigger length of \qty{2}{ms}, the resulting minimum acquisition length is \qty{2.5}{ms}. To reach the data-reduction goal of $\geq$\qty{96}{\percent}, the trigger rate has to be smaller than \qty{16}{Hz}. We tested thresholds above \qty{14}{\sig} ($32\,\si{Hz}$ simulated noise trigger rate) for the single-trigger configuration. For the trigger configurations with $\textnormal{n-triggers}>1$, thresholds as low as \qty{13}{\sig} were tested because the re-trigger requirement provides additional data reduction.

\begin{figure}
\centering
\includegraphics[width=\linewidth]{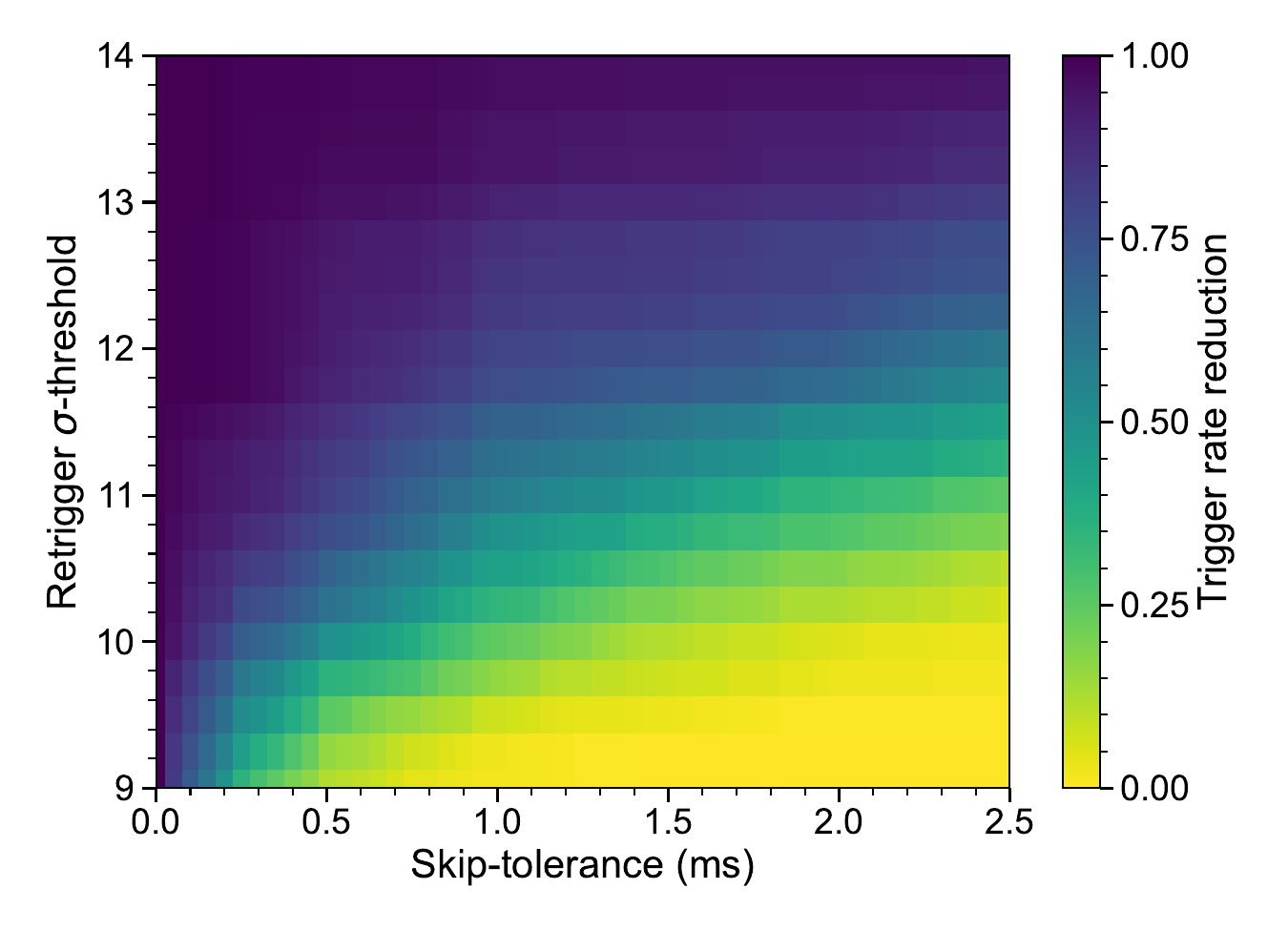}
\caption{Noise trigger rate reduction in simulated data by requiring a second FMT-trigger before starting an acquisition ($\textnormal{n-triggers}=2$). The color encodes the fraction of acquisitions that are not recorded by requiring that a second trigger-flag from a re-trigger threshold \unit{\sig} occurs within the time limit set by the skip-tolerance. The first trigger was set at \qty{13}{\sig}. 
}
\label{fig:ntriggers_2_data_reduction}
\end{figure}

For the n-triggers parameter, only two settings were tested. Since most events at Phase~II pressures are short ($<$\qty{1}{ms}) and hence not expected to exceed the trigger threshold very often in their lifetime, n-triggers was either set to 1 or 2. The additional data reduction achievable by using a second trigger ($\textnormal{n-trigger} = 2$) can be seen in \autoref{fig:ntriggers_2_data_reduction}.  
Based on this simulation, an FMT threshold range between \qty{10}{\sig} and \qty{13}{\sig} was tested. 
Below \qty{10}{\sig} the additional data reduction is not sufficient to achieve the data volume target. 

\subsubsection{Parallel trigger tests with \kr{} data}
\label{sec:parallel_trigger_tests}

\begin{figure}
\centering
\includegraphics[width=\linewidth]{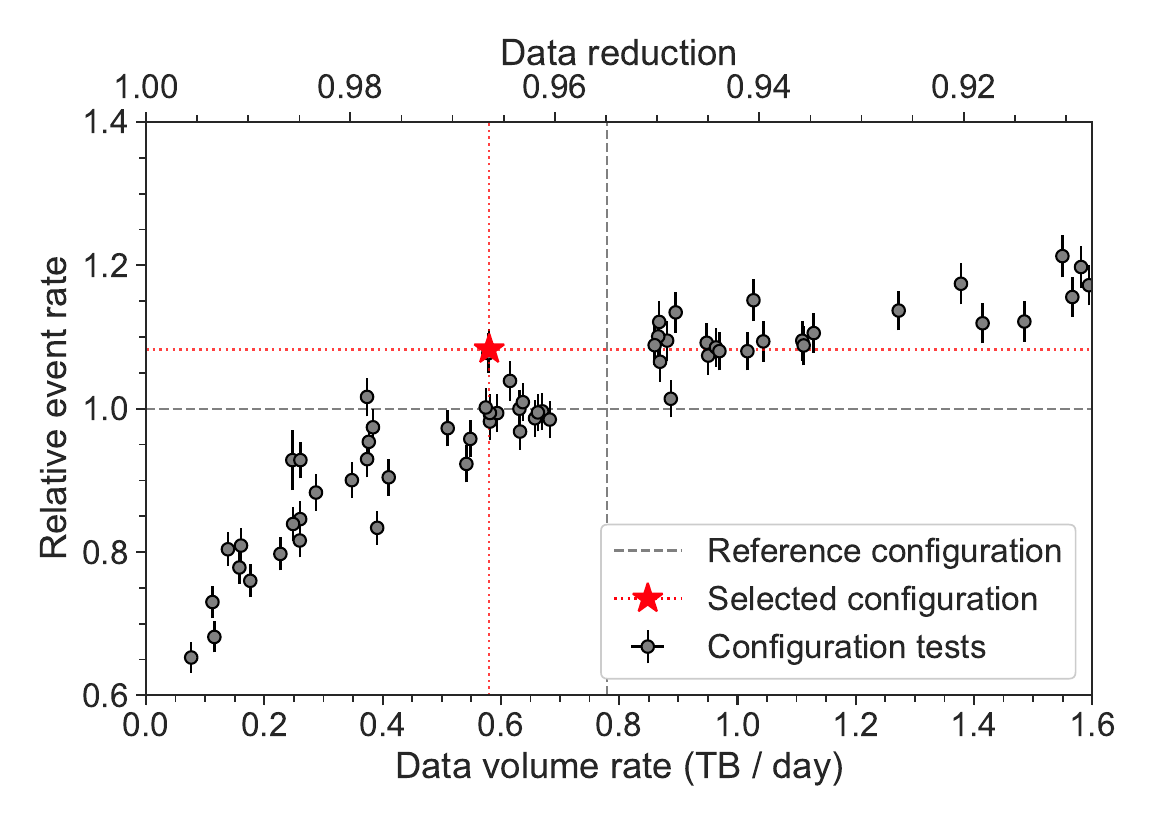}
\caption{Event rate vs. data volume in the trigger-optimization test runs. Each data point originates from a different trigger configuration. The horizontal error bars are smaller than the markers. The configuration marked by a red star was chosen for production data-taking. It achieves a detection efficiency improvement of 8\% over the reference configuration while reducing the data volume on disk by an additional 26\%. See \autoref{sec:parallel_trigger_tests} for a description of how the reference configuration was used.  With 96.7\% absolute data reduction the selected configuration satisfies the triggering requirement.
}
\label{fig:trigger_optimization}
\end{figure}

The trigger parameter optimization in Phase~II was done by maximizing the event rate in real data. Simulations were then used to study the performance of the trigger with respect to known true event properties (\autoref{sec:trigger_in_simulations}).
To find the optimum trigger configuration, \kr{} K-line electrons were recorded with all three ROACH2 channels, \textit{a}, \textit{b}, and \textit{c}, in parallel, with all 3 central frequencies of the down-converted 100-MHz band set to $\mathrm{IF}=\qty{1.41}{GHz}$ (the frequency location of the K-line events in the IF band). The Psyllid instance processing packets from channel \textit{c} served as reference and was always configured with the same trigger settings (in a configuration that had been found to work reasonably well), while the channel \textit{a} and \textit{b} Psyllid instances were used to test new configurations in each run.
This way, two configurations could be tested per parallel three-channel run. The reference channel is needed to ensure the result is not falsified by fluctuations of the krypton source activity over time.
The trigger parameters were scanned over the pre-defined parameter space (see \autoref{sec:trigger_parameter_space}), and the event rates after triggering and reconstruction were compared to obtain the relative efficiency and data reduction for each parameter combination.
\autoref{fig:trigger_optimization} shows the detected event rate versus the recorded data volume for all tested trigger configurations. The selected configuration (marked in red) for production data-taking achieves a data reduction of \qty{96.7}{\percent}. Note that for the Phase~II tritium data set \cite{phase_II_prl, phase_II_prc}, the cut thresholds in the offline reconstruction were higher to achieve a zero background rate. As a result the relative impact of the trigger efficiency on the detected event rate is sub-dominant since more low-SNR events are missed by the offline reconstruction. The efficiency of the trigger and the offline reconstruction under Phase~II tritium running conditions are compared in \autoref{sec:simulation_studies}.

\subsection{Trigger performance in simulation}
\label{sec:trigger_in_simulations}

In this section we show the performance of the optimized trigger configuration. 
The software trigger is tested using simulated events with signal properties (e.g. SNR, track and event duration) and ADC settings that mimic real data.
The simulated data are generated using the Locust package \cite{locust}. 
To imitate CRES data, Locust is configured to generate white noise and add CRES-like chirps to the time series. 
The simulated CRES-signal properties are randomly drawn from probability density functions describing their distributions, as described below, and the resulting chirp is added by Locust's \href{https://github.com/project8/locust_mc/blob/master/Source/Generators/LMCFakeTrackSignalGenerator.cc}{\textit{LMCFakeTrackSignalGenerator}}. The distributions of track duration and number of tracks in an event are found by doing an exponential and geometric fit respectively to the reconstructed real data. The SNR distribution is obtained from the calculated power coupled to a waveguide by simulated electrons in a magnetic trap (see, for example, \autoref{fig:trigger_snr_cut}).
It was shown that this way, the signal of real CRES electrons is well reproduced \cite{ter_paper}. 

\begin{figure}
\centering
\includegraphics[width=\linewidth]{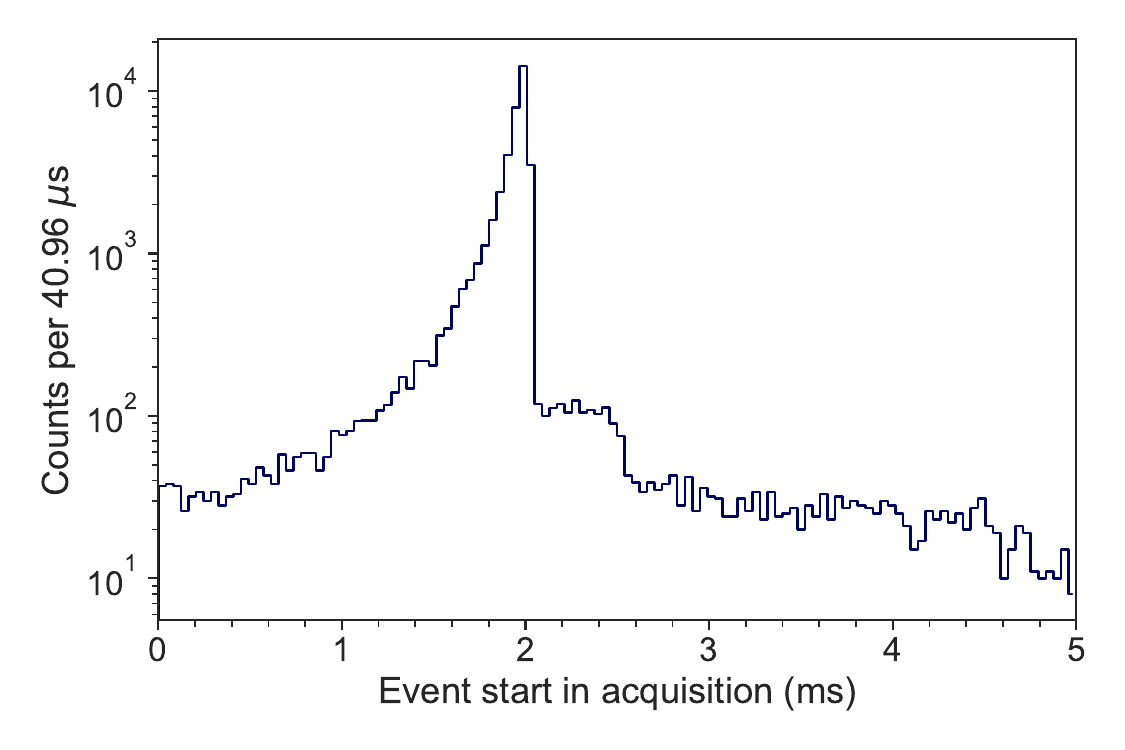}
\caption[Start times in triggered fake event acquisitions]{True start times of simulated events in triggered acquisitions. The pre-trigger time was set to \qty{2}{ms}, the skip-tolerance to \qty{0.5}{ms}. Only events that start within the acquisition are shown. Events starting before \qty{2}{ms} only triggered after the true event start. This exhibits the importance of the pre-trigger time. The plateau from \qtyrange{2}{2.5}{ms} originates from events that start during an ongoing acquisition initiated by a noise trigger.}
\label{fig:true_start_time_in_acq}
\end{figure}

Locust writes the generated time series to Egg files. In addition, the \textit{LMCFakeTrackSignalGenerator} writes the simulated event truths to a ROOT file. 
To process the simulated Egg files, a frequency-transform node is added to the Psyllid stream to read in and Fourier transform the simulated time-series data before processing it with the FMT and event-builder nodes. After processing, the read-in and triggered time-domain data are written to a second Egg file. The acquisition start and end times are logged in the file header, which allows for analysis of the true event-start times relative to the time of triggering (\autoref{fig:true_start_time_in_acq}).

\subsubsection{Trigger efficiency dependence on SNR and event duration}

\begin{figure}
\centering
\includegraphics[width=\linewidth]{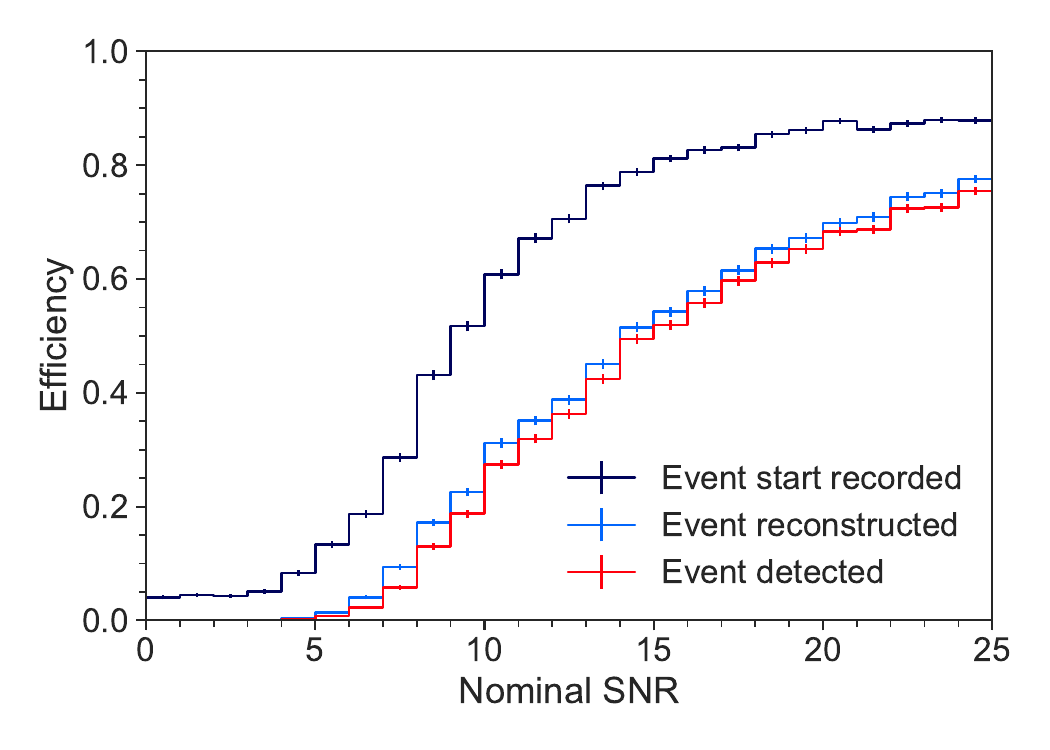}
\caption{Fraction of simulated events that are recorded by Psyllid (dark blue) or reconstructed offline (bright blue) as a function of the event SNR. The offline reconstruction is tuned to remove low-SNR or short events that cannot be distinguished from noise. Therefore, the combined detection efficiency (red) closely follows the offline reconstruction efficiency, and only a small additional fraction of otherwise reconstructed events go undetected when recording data with Psyllid. 
}
\label{fig:trigger_vs_reconstruction_efficiency}
\label{fig:event_recording_efficiency}
\end{figure}

\begin{figure}
\centering
\includegraphics[width=\linewidth]{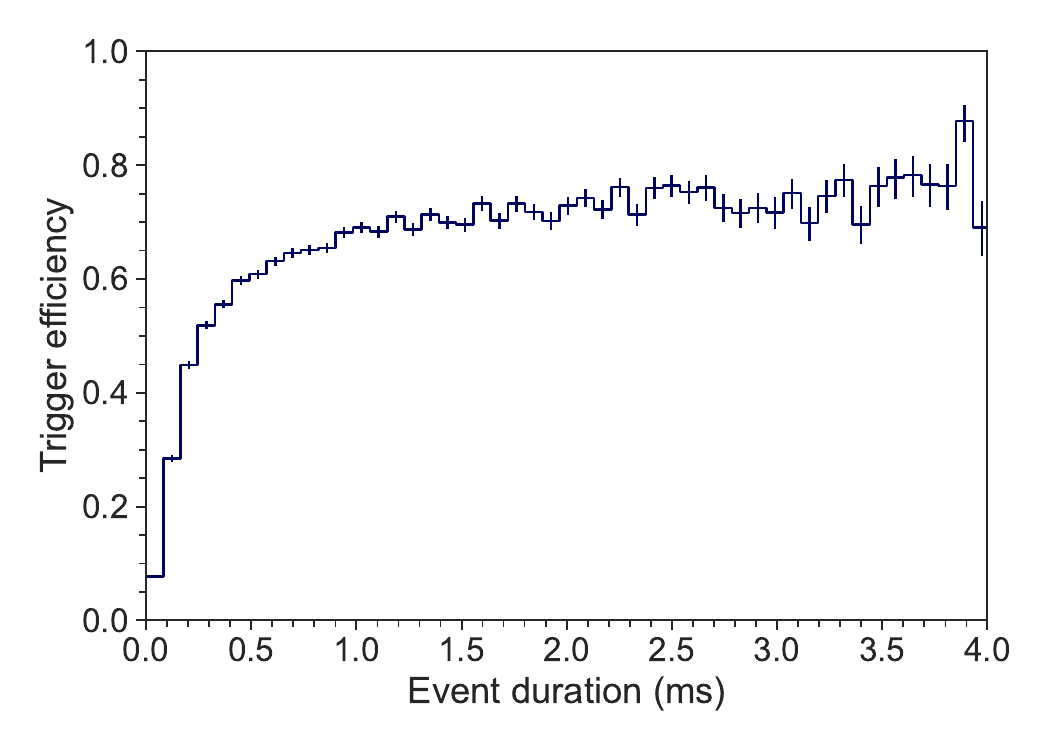}
\caption[Trigger efficiency dependence on event duration]{Dependence of the fraction of recorded simulated events on the event duration. Events were  simulated with random nominal SNR following a uniform distribution (range $0-25$). Because of noise fluctuations and the variation of event SNR in a spectrogram (caused by the upward-sloped tracks in conjunction with the frequency response of the DFT), longer events are more likely to surpass the trigger thresholds during their lifetime.}
\label{fig:length_trigger_efficiency}
\end{figure}

In a CRES experiment, the most important data point of each event is the start of the signal. In the following analysis, an event is only counted as detected if the event start was recorded. To measure the efficiency vs. event SNR, simulated events of various SNRs were generated and processed by Psyllid in triggered mode. The efficiency is calculated as the fraction of events that started within the acquisition. The trigger efficiency dependence on simulated SNR is shown in \autoref{fig:event_recording_efficiency}.
In the low-SNR limit, the recording efficiency corresponds to the fraction of recorded time resulting from noise triggers. This is because in this SNR regime the events themselves do not lead to sufficient received power in order to trigger an acquisition. Only for SNR $\geq 3$ does the combination of noise power and signal power start to lead to an increase of recorded events above the random noise trigger efficiency. Because event duration follows an exponential distribution, $\approx$\qty{10}{\percent} of events are shorter than one \qty{40.96}{\us} record length (the exact fraction varied depending on run conditions); their power is likely to be split between two records with diminished power in each. With {n-triggers}~$=2$, events that are not longer than one record are only likely to be detected if a noise fluctuation causes the second trigger. Hence the trigger efficiency does not converge to 1 for high SNR.
A further dependence of the trigger efficiency on the event duration results from longer events having a higher chance of overcoming the trigger thresholds (\autoref{fig:length_trigger_efficiency}).


\subsubsection{Comparison of trigger and reconstruction efficiencies}\label{sec:simulation_studies}

\begin{figure}
\centering
\includegraphics[width=\linewidth]{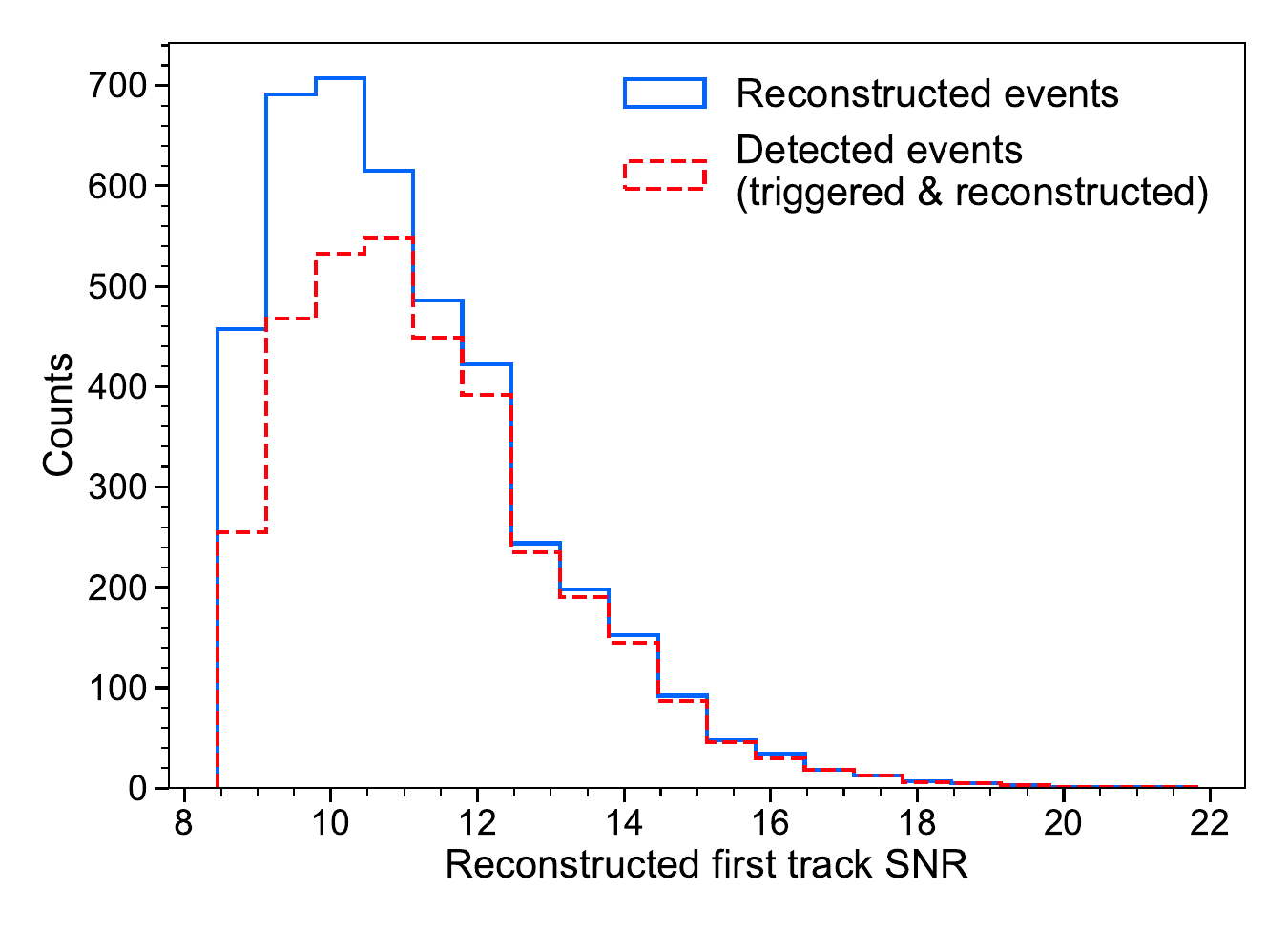}
\caption{Histogram of reconstructed SNR from a simulated data set that mimics Phase~II data. The simulated SNR distribution of all events (not shown) resembles an exponential distribution. When both the triggering and the reconstruction are applied (dashed red), 18\% of the events that could have been found by the offline reconstruction alone (solid blue) are removed.}
\label{fig:trigger_snr_cut}
\end{figure}

After recording by Psyllid, Egg files are transferred to a computing cluster for offline event reconstruction. For tritium data-taking, the offline reconstruction was optimized for efficiency (event rate) while maintaining a false event rate below 1 per \qty{100}{days} per channel at \qty{90}{\percent} confidence \cite{ter_paper}. In contrast, the trigger was optimized for efficiency while observing the constraint on data volume. Here the reconstruction and trigger efficiency are compared for simulated events. In \autoref{fig:trigger_vs_reconstruction_efficiency} it can be seen that the event reconstruction is stricter than the trigger and does not tolerate the misidentification of noise fluctuations as events. Consequently the efficiency of reconstructing events is significantly lower than the efficiency of triggering on events. Conversely, for events of a uniform SNR distribution, only a small fraction of events that could have been reconstructed are removed by the trigger.
Under Phase~II data-taking conditions, the majority of detected events have a reconstructed SNR between 8 and 11 (\autoref{fig:trigger_snr_cut}). In this range the relative impact of the trigger efficiency on the event detection (trigger plus offline reconstruction) is larger than at higher SNR (see \autoref{fig:trigger_vs_reconstruction_efficiency}). As a result, \qty{18}{\percent} of events that could be reconstructed offline are not recorded. Low-SNR events correspond to electrons with a large longitudinal trajectory in the magnetic trap and the waveguide. Due to the resulting frequency modulation \cite{ashtari2019a}, the power in the main carrier frequency is reduced. Because these electrons experience a different average magnetic field along their trajectory than other electrons, they are detected at a different frequency according to \autoref{eq:cres}. In addition, the event starting points are more likely to be blurred by noise and the frequencies reconstructed by the offline analysis are on average less precise \cite{ter_paper}. The trigger inefficiency for these events therefore narrows the spread of detected frequencies and improves the frequency resolution of recorded spectra.

\subsection{Trigger efficiency vs. frequency}

In the Project~8 Phase~II experiment, all three ROACH2 channels were used to record tritium data in parallel. The central frequency of each channel was set such that the channels covered different but overlapping frequency bands. This way, the upper \qty{2.6}{keV} of the tritium spectrum and \qty{1.1}{keV} past the endpoint were recorded. Tritium spectroscopy aiming for a neutrino mass measurement requires a precise shape analysis of the spectral endpoint region. The trigger efficiency as a function of frequency must therefore be precisely known. If this variation is not negligible compared to the statistical power of the recorded spectrum, it must be included in the spectral analysis. We measured the trigger efficiency across a single channel's frequency band and found a configuration of the central frequencies that keeps the frequency dependence of the trigger efficiency small enough that it can be ignored in the Phase~II tritium analysis.

\subsubsection{Trigger efficiency in a single channel}
\begin{figure}
\centering
\begin{subfigure}[t]{\linewidth}
\includegraphics[width=\textwidth]{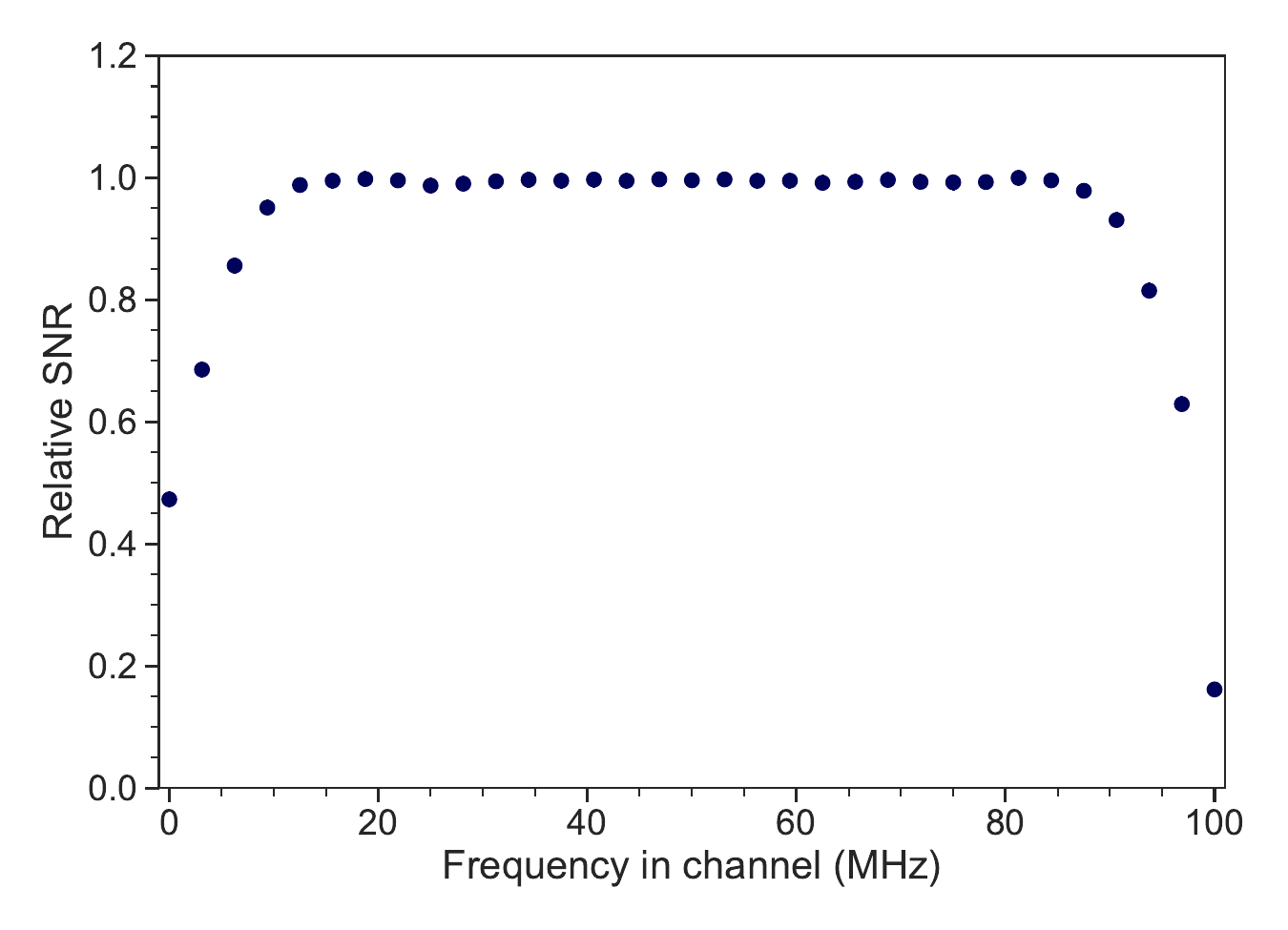}
\caption[ROACH2 channel SNR vs. frequency]{ } 
\label{fig:awg_snr_vs_frequency}
\end{subfigure}

\begin{subfigure}[t]{\linewidth}
\includegraphics[width=\textwidth]{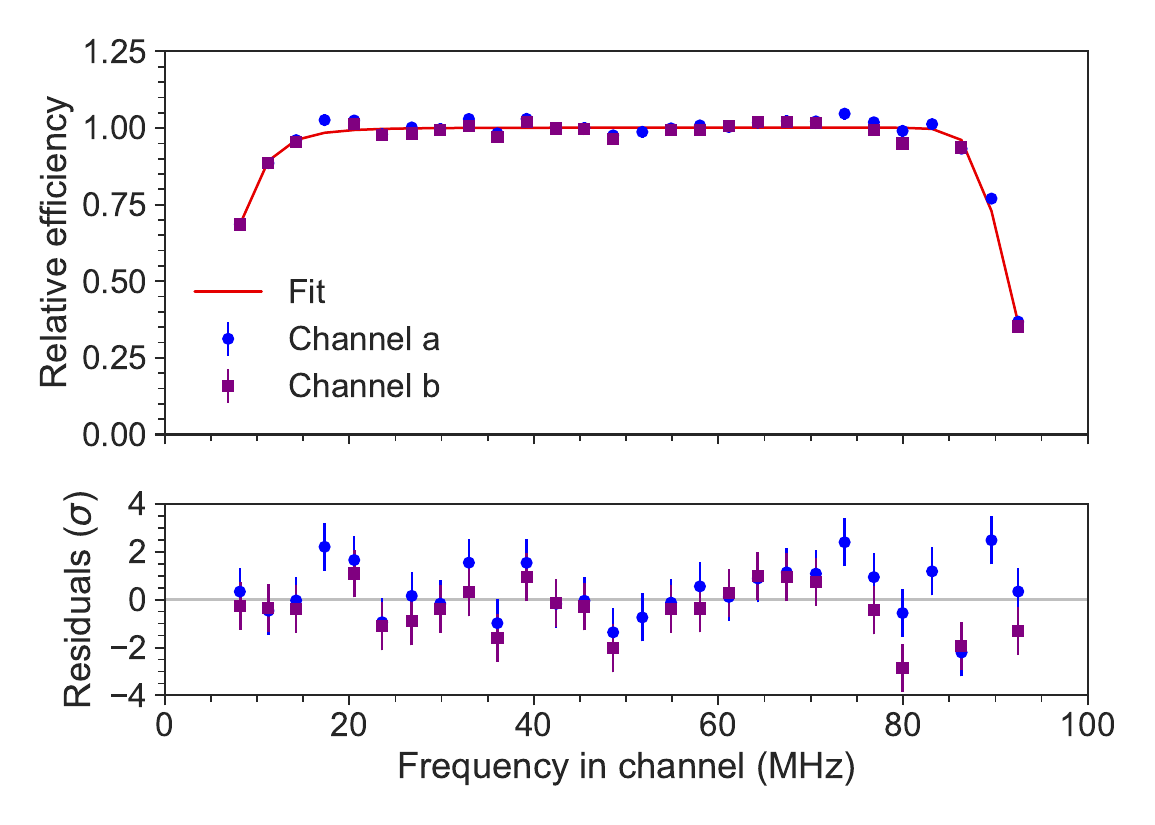}
\caption[Detection efficiency vs. frequency in the ROACH2 channels]{ } 
\label{fig:cf_scan_channel_efficiency}
\label{fig:cf_scan_channel_efficiency_fitted}
\end{subfigure}
\caption[SNR and trigger efficiency in ROACH2 channels]{SNR (a) and trigger efficiency (b) as a function of frequency in the ROACH2 channels. The SNR was measured by injecting a test tone, whereas the efficiency was measured with \kr{} data. The trigger efficiency dependence on frequency originates from the SNR roll-off at the channel edges as a result of noise aliasing. The trigger efficiency is further reduced at high frequencies by the detection inefficiency for short tracks (at the upper edge of the frequency band, tracks are cut short because of their upward slope). The efficiency is fitted with \autoref{eq:filter_function}. The fitted parameters are $A=1.001\pm0.002$, $f_{\textnormal{cut},1}=(89.8 \pm 0.1)\,\si{MHz}$, $f_{\textnormal{cut},2}=(8.4 \pm 0.1)\,\si{MHz}$, $p_1 = 31.4 \pm 1.3$ and $p_2 = -2.3 \pm 0.2$.}
\end{figure}

\begin{figure}
\includegraphics[width=\linewidth]{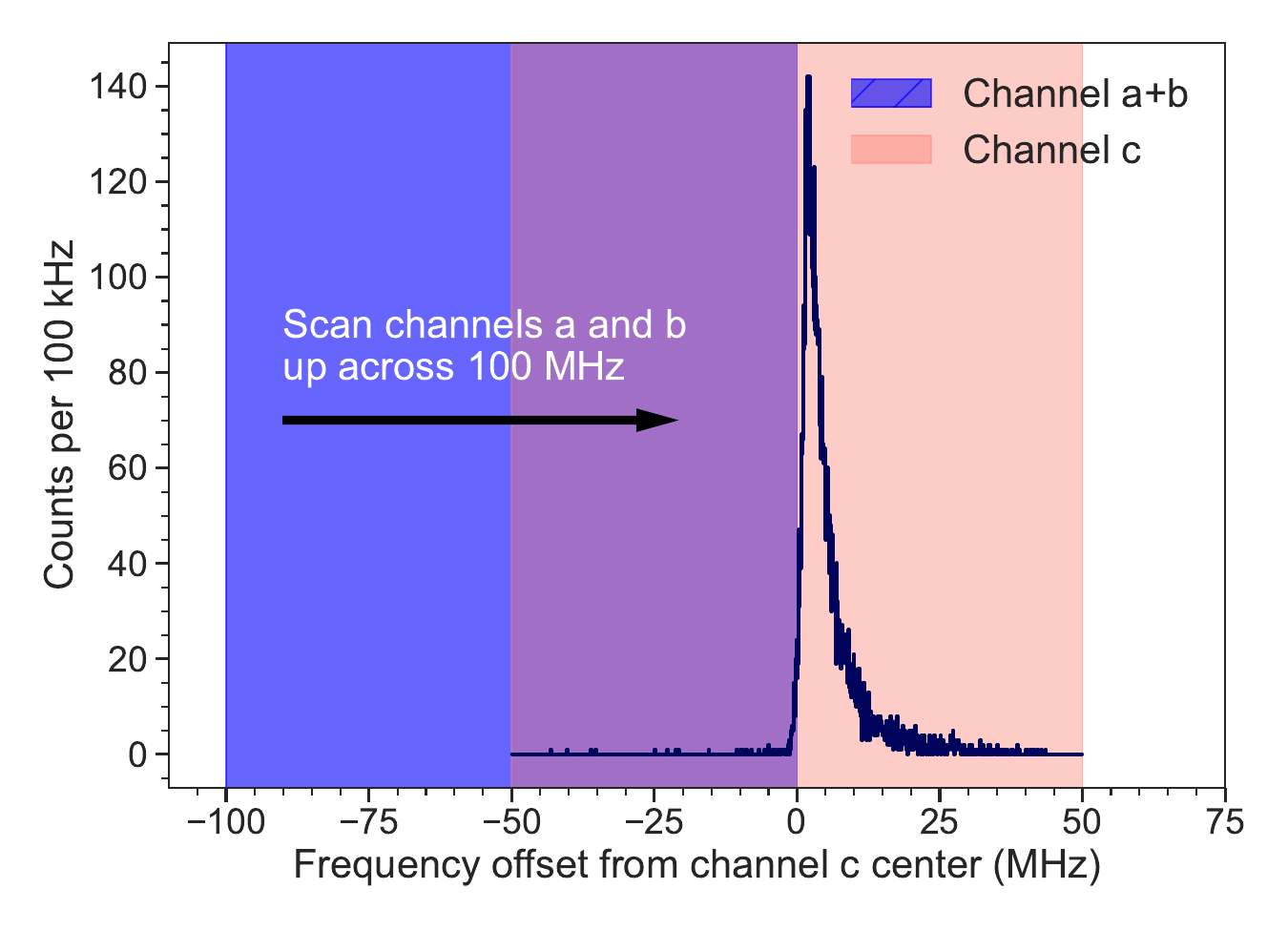}
\caption[Central frequency scan procedure]{The detection efficiency as a function of frequency is measured by comparing count rates in channels \textit{a} and \text{b} with the count rate in channel \textit{c}. The central frequencies of channels \textit{a} and \textit{b} (blue) are scanned in 3.125\,\si{MHz} steps across the frequency band of channel \textit{c} (light red) that is centered on the $\mathrm{^{83m}Kr}$ K-line frequency (the histogram is the measured K-line distribution). Dividing the rate of detected events in the overlapping region of a moving channel (purple) by the rate in the stationary reference channel \textit{c} yields the relative detection efficiency (see \autoref{fig:cf_scan_channel_efficiency_fitted}).}
\label{fig:channel_scanning}
\end{figure}

During the first down-conversion stage on the ROACH2, the signal is band-pass filtered. Due to aliasing of the remaining noise outside the Nyquist zone, the SNR of a signal decreases towards the edges of a channel's frequency band. To measure the SNR dependence on frequency in a channel, a tone of fixed frequency is injected while a single channel's central frequency is scanned from 50\,\si{MHz} below to 50\,\si{MHz} above the tone frequency. \autoref{fig:awg_snr_vs_frequency} shows the ratio of the recorded tone power to the noise floor as a function of the tone's frequency position in the channel. The event-detection efficiency is expected to follow a similar profile. To investigate the relative detection efficiency in a channel (\autoref{fig:cf_scan_channel_efficiency}), \kr{} K-line electrons were recorded using all three channels simultaneously. This time, the central frequencies of two channels were scanned from \qty{50}{MHz} below the \qty{17.8}{keV} peak position to \qty{50}{MHz} above. The third channel was kept centered on the peak and provided the reference count rate (\autoref{fig:channel_scanning}).

For the count rate to be indicative of the efficiency at the peak position in a channel, only events close to the peak center $f_0$ are counted: $|f -f_0| \leq \qty{1}{MHz}$.  Dividing the event rates in the two scanned channels by the rate in the reference channel yields the detection efficiency dependence on frequency relative to the central frequency in the reference channel.
The result is shown in \autoref{fig:cf_scan_channel_efficiency_fitted}. The efficiencies in the two channels agree and they are fitted together with a band-pass filter function of two independent orders, $p_1$ and $p_2$ and cut-off frequencies, $f_{\rm cut, 1}$ and $f_{\rm cut, 2}$:
\begin{equation}\label{eq:filter_function}
\textnormal{Gain}(f) = A\cdot 
\frac{1}
{\sqrt{1+ \left( \frac{f}{f_{\textnormal{cut}, 1}} \right) ^{2 p_1}}} \cdot 
\frac{1}
{\sqrt{1+ \left( \frac{f}{f_{\textnormal{cut}, 2}} \right) ^{2 p_2}}}.
\end{equation}
$A$ is a frequency-independent scaling of the gain.  The cut-off frequencies and orders are independent fit parameters because the efficiency roll-off is asymmetric. This results from the fact that events at high frequencies can drift out of the channel's frequency band and are cut short in the reconstruction. Because of the dependence of the trigger and reconstruction efficiency on event duration, the detection efficiency decreases more sharply at the upper frequency edge of the channel.

\subsubsection{Channel configuration for Phase~II}\label{sec:tritium_channel_configuration}

\begin{figure}
\centering
\begin{subfigure}[t]{\linewidth}
\includegraphics[width=\textwidth]{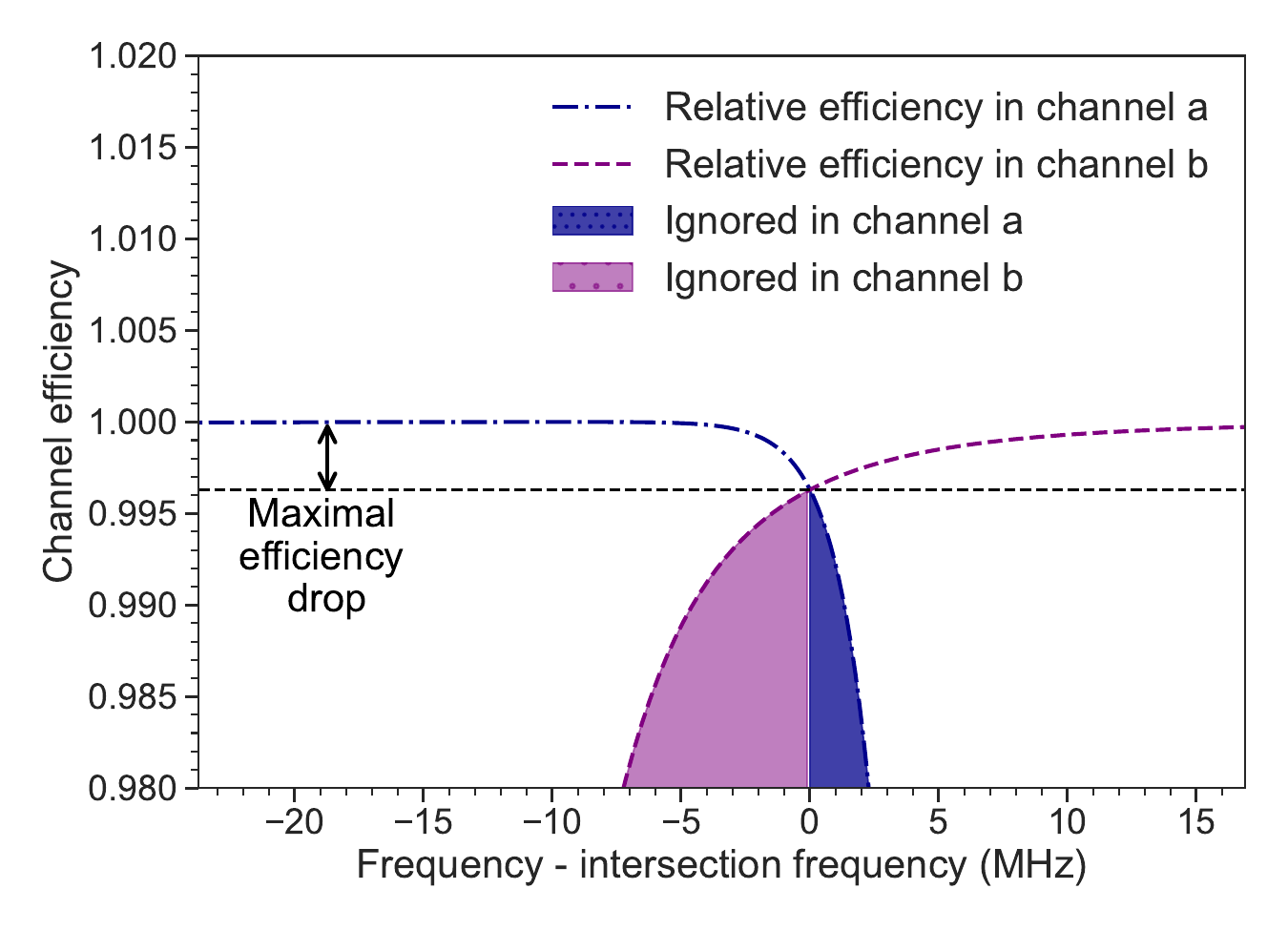}
\caption{ }
\label{fig:channel_overlap_drawing}
\end{subfigure}

\begin{subfigure}[t]{\linewidth}
\includegraphics[width=\textwidth]{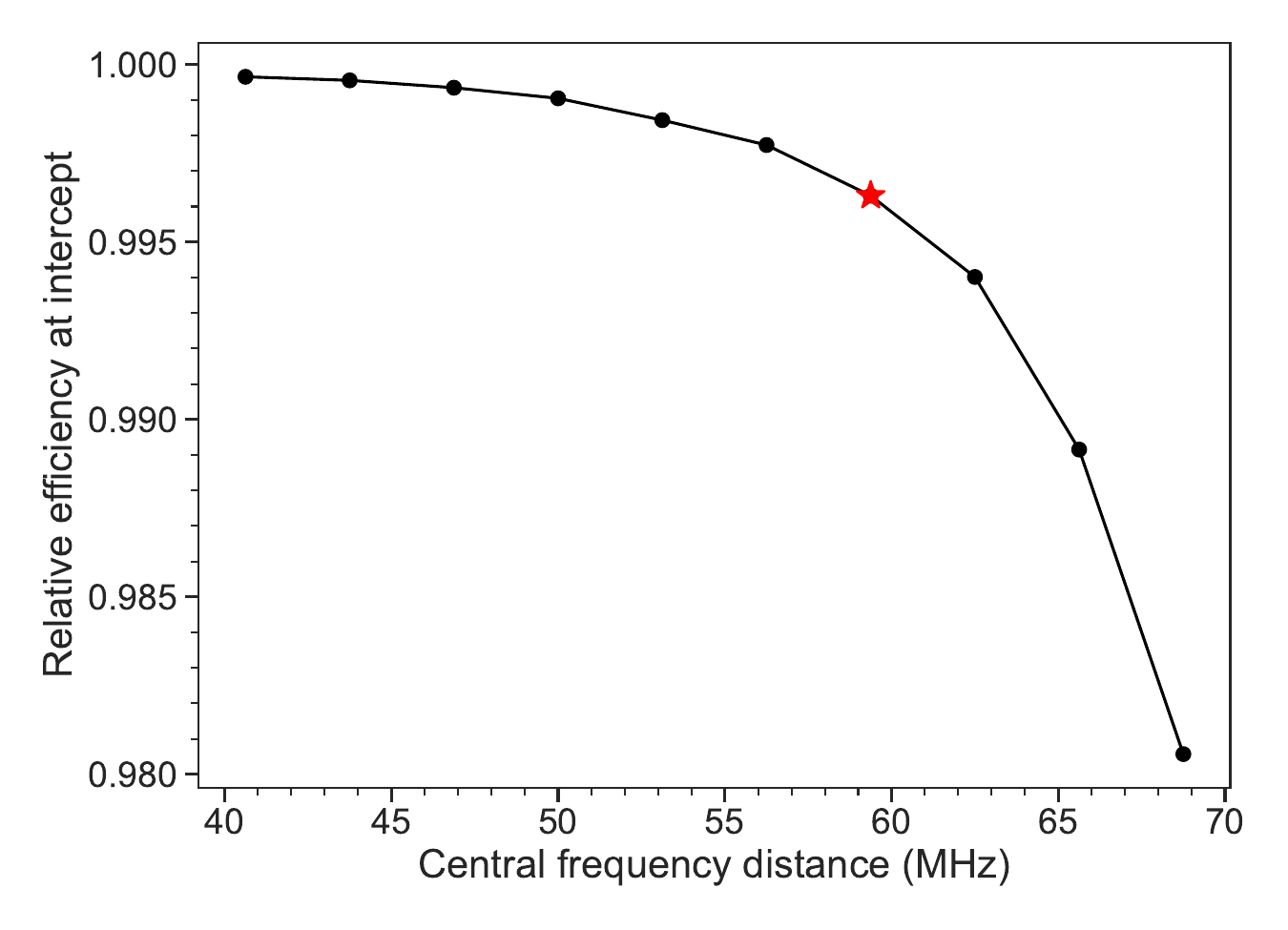}
\caption{ }
\label{fig:efficiency_drop}
\end{subfigure}
\caption{Efficiency drop between two channels: When channels are placed adjacent to one another and their frequency bands overlap, the channel efficiencies intersect (a). Events recorded in the shaded regions are ignored in the analysis to prevent double counting. The magnitude of the efficiency at the intersection point (b) depends on the distance between the two central frequencies. The red star in (b) marks the selected channel configuration for Phase II tritium data-taking.}
\label{fig:cf_selection_process}
\end{figure}

\begin{figure}
\centering
\includegraphics[width=\linewidth]{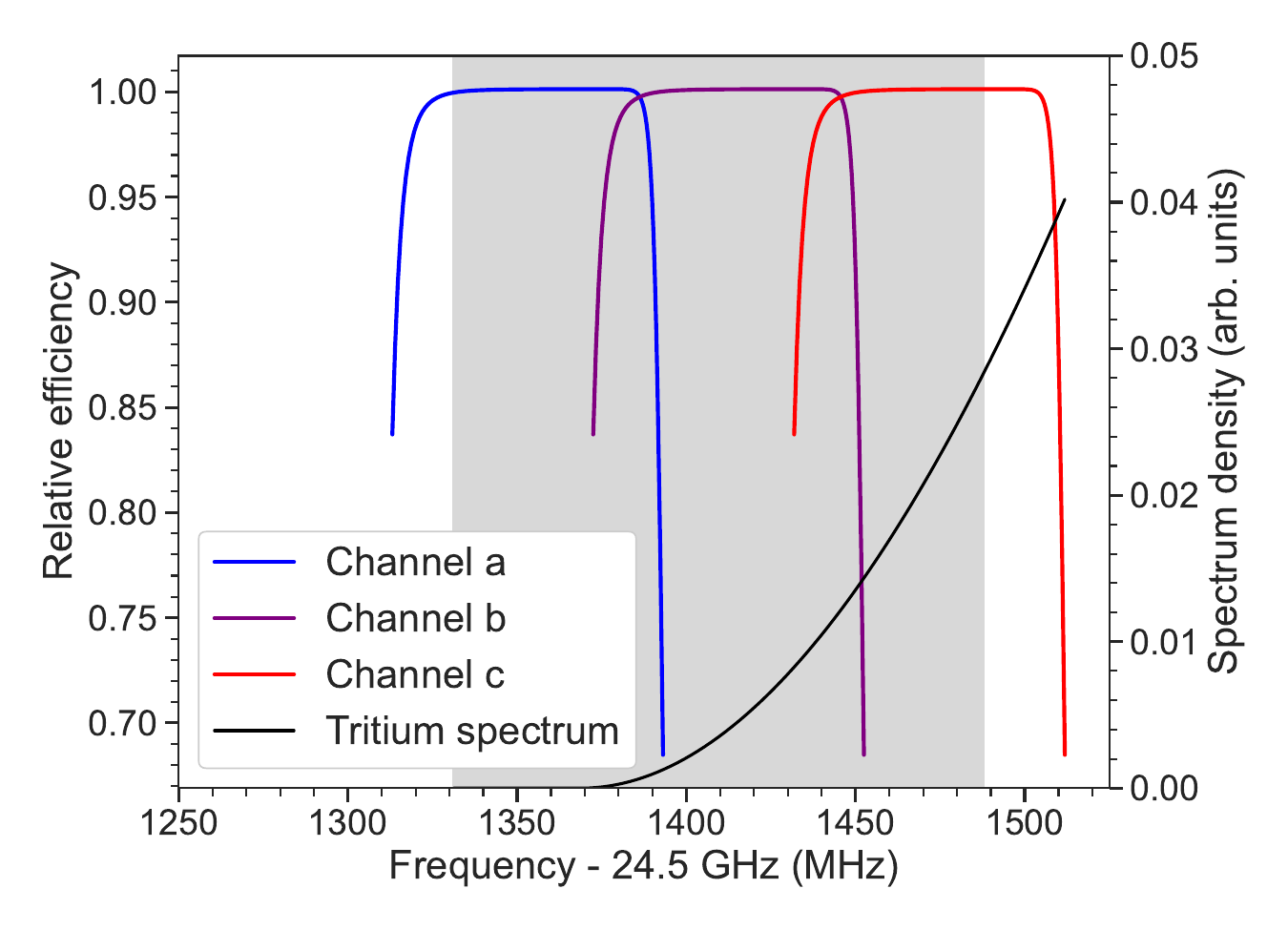}
\caption{Channel configuration for tritium data-taking, overlayed by the tritium spectrum. In this configuration the efficiency drop between channels is $< \qty{0.5}{\percent}$ and could therefore be neglected in the Phase~II tritium analysis \cite{phase_II_prc}. The grey shaded area marks the tritium analysis region of interest (corresponding to the energy range \qty{16.2}{}--\qty{19.4}{keV}).}
\label{fig:tritium_channel_configuration}
\end{figure}

Using the asymmetric fitted filter function from \autoref{fig:cf_scan_channel_efficiency_fitted}, the maximum efficiency loss for any given distance between two channel central frequencies
can be calculated as illustrated in \autoref{fig:cf_selection_process}. Note that the distance between the central frequencies in \autoref{fig:efficiency_drop} takes discrete values since the ROACH2 central frequencies can only be set in steps of \qty{3.125}{MHz}. For Phase~II tritium data-taking, we required that the maximum efficiency decrease be $<\qty{0.5}{\percent}$ (\autoref{fig:tritium_channel_configuration}).  An efficiency variation on this scale is negligible compared to the statistical uncertainty of the recorded tritium spectrum and it was hence neglected in the data analysis \cite{phase_II_prl, phase_II_prc}. Events recorded in a channel beyond the intersection point with the neighboring channel were cut during data analysis to avoid double counting.  The central frequencies for each channel and the transition frequencies between channels are shown in Table~\ref{tab:freq_params}. For context,  Table~\ref{tab:freq_params} also provides the definition of the frequency window used in the analysis of the tritium spectrum (shaded region in \autoref{fig:tritium_channel_configuration}) and the frequency of the T$_2$ beta-decay endpoint.

\begin{table}
    \centering
     \begin{tabular}{lS[table-format=3.2]}
        \toprule
        Parameter & \parbox{2.1cm}{\centering IF Frequency (GHz)} \\
        \midrule
        Minimum T$_2$ analysis freq. & 1.33 \\
        Center frequency, channel a & 1.353 \\ 
        Transition freq., channels a$\to$b & 1.386 \\ 
        T$_2$ endpoint & 1.371 \\
        \kr{} K-line & 1.41 \\
        Center frequency, channel b & 1.413 \\ 
        Transition freq., channels b$\to$c & 1.446 \\ 
        Center frequency, channel c & 1.472 \\ 
        Maximum T$_2$ analysis freq. & 1.49 \\
        \bottomrule
    \end{tabular}
    \caption{Frequency parameters for the three channels as used for Project~8, Phase~II. 
 The minimum and maximum frequencies for the analysis of the tritium spectrum (shaded region in \autoref{fig:tritium_channel_configuration}), the T$_2$ endpoint frequency, and the \kr{} K-line frequency are given as context. The full frequency range corresponds to the calibration range of the detector response \cite{phase_II_prc} and is not limited by the DAQ.}
    \label{tab:freq_params}
\end{table}

\subsection{Summary}
The DAQ system described in \autoref{sec:system_description} has been proven to be suitable for recording CRES events in Phase~II. By optimizing the configurable trigger parameters, a data reduction of \qty{96.7}{\percent} was achieved. The running conditions for Phase~II tritium data-taking were chosen to maximize the event rate by increasing the gas pressure and using a deep magnetic trap. Under these conditions, the trigger misses \qty{18}{\percent} of all events that could be reconstructed offline. The majority of these events are of low SNR and offline reconstruction would yield an increased uncertainty on their start frequency \cite{ter_paper}.

The trigger efficiency is uniform across the center of a channel's frequency band but decreases towards the edges due to noise aliasing. To record a wide frequency range with all three channels in parallel, the central frequencies were chosen such that the efficiency drop between channels was small enough to be ignored in Phase~II. The DAQ therefore could be eliminated as a source of systematic uncertainty in the analysis of the tritium spectrum \cite{phase_II_prl, phase_II_prc}. 

\section{Outlook for future CRES experiments}
\label{sec:outlook}
A primary goal of the next phase of Project 8 (Phase~III) is to demonstrate the use of CRES for measurements using a large-volume source. The design currently being pursued by the Project~8 collaboration uses a large resonant cavity to detect a narrow range of signal frequencies, as described in~\cite{Project8:Snowmass2022}. 
Project~8 plans to demonstrate the feasibility of this technique using a series of demonstrator experiments that will eventually have resonant frequencies of \qty{\leq 1}{GHz}.
At this frequency, regardless of the width of the final frequency band, the data can be directly digitized using modern signal-processing hardware.  We are currently evaluating the AMD Zynq Ultrascale+ ZCU216 RF System-on-Chip (RFSoC) platform, but other hardware may be chosen for the final system.  This type of system has the advantage of providing a highly integrated platform that includes high-rate digitization, a powerful FPGA for signal processing, and high-speed Ethernet connectivity.  Overall the capabilities are similar to those provided by the ROACH2 system, but in an off-the-shelf commercial package.  Signal-processing optimizations can also be made to avoid the efficiency rolloffs at the edges of the frequency window(s) caused by image noise.

The data acquisition for a cavity-based experiment will need to digitize a single or a few channels. The sampling rate is limited on the lower end by the size of the frequency band corresponding to the energy and frequency region of interest. The frequency range to be covered by electron signals depends on the variation of magnetic field experienced by each electron and on the frequency of each electron's axial motion in the magnetic trap. This range can then be mixed down to decrease the required data-storage volume. If no down sampling is performed, the sampling rate would be larger than \qty{2}{GS/s} by a margin that allows clean low-pass filtering of the signal. A \qty{2}{GS/s} sampling rate for simplicity, with IQ sampling at 8-bit, results in a data rate of \qty{2}{GB/s}, and a total volume of \qty{63}{PB} per channel for a 1-year runtime. This volume can be reduced by means of triggering and real-time signal reconstruction. 

In Phase~III the goal is to demonstrate the ability to handle event rates significantly higher than in Phase~II and eventually it is expected that events could be ongoing at any point in time. While the information of main interest for performing spectroscopy is the event-start frequencies, other features such as the full track durations, the number of tracks, and the distribution of frequency jumps between them contain important information about scattering and gas composition that is needed in the analysis of the spectrum~\cite{phase_II_prc}. Several options are being studied that would allow for a reduction in the overall data volume, while still providing the information necessary for precision spectroscopy.  These include storing raw data at the start of each event and reconstructed information about the remainder of the event, and storing a narrow band of frequencies around the signal.  For any data-reduction option, random triggers that record periods of raw data independent of information about ongoing events would allow for the extraction of unbiased statistical information about the track properties.

\section{Conclusion}

In this work we have described the data-acquisition system for Phase~II of Project~8, in which we demonstrated the ability to record three channels each with a \qty{100}{MHz}-wide bandwidth in parallel and perform real-time signal processing and triggering on the data to enable efficient data storage at the relevant event rate.  We studied the signal and noise properties such that the systematic uncertainties for the CRES analysis could produce the first precision measurement of the tritium beta-decay endpoint and neutrino mass limit with CRES \cite{phase_II_prl, phase_II_prc}.  The lessons learned will inform the design and implementation of the data-acquisition system for future phases of Project~8.




\section{Acronyms}
\begin{description}
\item[ADC] Analog-to-Digital Converter.
\item[CASPER] Collaboration for Astronomical Signal Processing and Electronics Research.
\item[CRES] Cyclotron Radiation Emission Spectroscopy.
\item[DAQ] Data Acquisition.
\item[FIR] Finite Impulse Response.
\item[FPGA] Field-Programmable Gate Array.
\item[HDF5] Hierarchical Data Format, version 5.
\item[IF] Intermediate Frequency.
\item[IIR] Infinite Impulse Response.
\item[IQ] In-phase / Quadrature.
\item[RF] Radio-Frequency.
\item[ROACH2] Reconfigurable Open-Architecture Computing Hardware version 2.
\item[SNR] Signal-to-Noise Ratio.
\item[UDP] User Datagram Protocol.
\end{description}

\section{Acknowledgments}

This material is based upon work supported by the following sources: the U.S. Department of Energy Office of Science, Office of Nuclear Physics, under Award No.~DE-SC0020433 to Case Western Reserve University (CWRU), under Award No.~DE-SC0011091 to the Massachusetts Institute of Technology (MIT), under Field Work Proposal Number 73006 at the Pacific Northwest National Laboratory (PNNL), a multiprogram national laboratory operated by Battelle for the U.S. Department of Energy under Contract No.~DE-AC05-76RL01830, under Early Career Award No.~DE-SC0019088 to Pennsylvania State University, under Award No.~DE-FG02-97ER41020 to the University of Washington, and under Award No.~DE-SC0012654 to Yale University; the National Science Foundation under grant No.~PHY-2209530 to Indiana University, and under grant No.~PHY-2110569 to MIT; the Cluster of Excellence “Precision Physics, Fundamental Interactions, and Structure of Matter” (PRISMA+ EXC 2118/1) funded by the German Research Foundation (DFG) within the German Excellence Strategy (Project ID 39083149); the Karlsruhe Institute of Technology (KIT) Center Elementary Particle and Astroparticle Physics (KCETA); Laboratory Directed Research and Development (LDRD) 18-ERD-028 and 20-LW-056 at Lawrence Livermore National Laboratory (LLNL), prepared by LLNL under Contract DE-AC52-07NA27344, LLNL-JRNL-847024; the LDRD Program at PNNL; and Yale University.  Portions of the research were performed using Research Computing at PNNL and the HPC cluster at the Yale Center for Research Computing.  The $^{83}$Rb/$^{83{\rm m}}$Kr isotope used in this research was supplied by the United States Department of Energy Office of Science through the Isotope Program in the Office of Nuclear Physics.

\section{Author contributions}

\noindent \textbf{C. Claessens}: Software, Validation, Formal Analysis, Investigation, and Writing - Original Draft; \textbf{N. S. Oblath}: Conceptualization, Software, Validation, Investigation, and Writing - Original Draft; \textbf{A. Young}: Methodology, Software, Validation, Writing - Original Draft; \textbf{B. H. LaRoque}: Software, Investigation, Writing - Review \& Editing; \textbf{A. Ashtari Esfahani, N. Buzinsky, R. Cervantes, L. de Viveiros, M. Fertl, L. Gladstone, M. Guigue, A. Lindman, E. Novitski, W. Pettus, L. Saldaña, P. L. Slocum, Y.-H. Sun, P. T. Surukuchi, M. Thomas, L. Tvrznikova, T. E. Weiss, E. Zaya, }and\textbf{ A. Ziegler: }Investigation and Writing - Review \& Editing; \textbf{A. Banducci, M. Grando, J. Hartse, A. M. Jones, J. A. Nikkel, A. B. Telles, }and\textbf{ F. Thomas: }Writing - Review \& Editing; \textbf{S. Böser: }Supervision, Funding Acquisition, and Writing - Review \& Editing; \textbf{B. Monreal, J. A. Formaggio, K. M. Heeger, K. Kazkaz, R. G. H. Robertson, G. Rybka, T. Thümmler, }and\textbf{ B. A. VanDevender: }Funding Acquisition and Writing - Review \& Editing

\bibliographystyle{elsarticle-num}
\bibliography{roach2}


\end{document}